\documentclass{aa}  
\usepackage{graphicx}
\usepackage{txfonts}
\usepackage{makecell}

\newcommand{\Prot}{P_{\rm{rot}}}
\newcommand{\Pip}{P_{\rm{IP}}} 
\newcommand{\hip}{h_{\rm{IP}}} 
\newcommand{\Teff}{T_{\rm{eff}}}
\newcommand{\logg}{log\,g}
\newcommand{\Rvar}{R_{\rm var}}
\newcommand{\nsel}{142,168 } 
\newcommand{\nseltwo}{141,151 } 
\newcommand{\nper}{67,515 } 
\newcommand{\newper}{20,397 } 
\newcommand{\ntot}{67,163 }

\defcitealias{McQuillan2014}{McQ14}
\defcitealias{Berger2020}{B20}
\defcitealias{Santos2021}{S21}

\begin{document} 

\let\oldpageref\pageref
\renewcommand{\pageref}{\oldpageref*}

    \title{New rotation period measurements of \ntot \textit{Kepler} stars\thanks{Table~\ref{period_table} is only available in electronic form at the CDS via anonymous ftp to cdsarc.cds.unistra.fr (130.79.128.5) or via https://cdsarc.cds.unistra.fr/cgi-bin/qcat?J/A+A/.}}

    \author{Timo Reinhold \inst{1}
        \and Alexander I. Shapiro \inst{1}
        \and Sami K. Solanki \inst{1}
        \and Gibor Basri \inst{2}
        }

    \institute{Max-Planck-Institut f\"ur Sonnensystemforschung, Justus-von-Liebig-Weg 3, 37077 G\"ottingen, Germany
    \email{reinhold@mps.mpg.de}
    \and Department of Astronomy, University of California, Berkeley, CA 94720, USA}
    \date{\today}

 
\abstract
{The \textit{Kepler} space telescope leaves a legacy of tens of thousands of stellar rotation period measurements. While many of these stars show strong periodicity, there exists an even bigger fraction of stars with irregular variability for which rotation periods are rarely visible or in most cases unknown. As a consequence, many stellar activity studies might be strongly biased toward the behavior of more active stars, for which rotation periods have been determined.}
{To at least partially lift this bias, we apply a new method capable of determining rotation periods of stars with irregular light curve variability. This effort greatly increases the number of stars with well-determined periods, especially for stars with small variabilities similar to that of the Sun.}
{To achieve this goal, we employ a novel method based on the Gradient of the Power Spectrum (GPS). The maximum of the gradient corresponds to the position of the inflection point (IP), i.e., the point where the curvature of the high-frequency tail of the power spectrum changes its sign. It was shown previously that the stellar rotation period $\Prot$ is linked to the inflection point period $P_{\rm IP}$ by the simple equation $\Prot = P_{\rm IP}/\alpha$, where $\alpha$ is a calibration factor. The GPS method is superior to classical methods (such as auto-correlation functions (ACF)) because it does not require a repeatable variability pattern in the time series, making it an ideal tool for detecting periods of stars with very short-lived spots.}
{From the initial sample of \nsel stars with effective temperatures $\Teff\leq6500\,K$ and $\logg\geq4.0$ in the \textit{Kepler} archive, we could measure rotation periods for \ntot stars by combining the GPS and the ACF method. We further report the first determination of a rotation period for \newper stars. The GPS periods show good agreement with previous period measurements using classical methods, where these are available. Furthermore, we show that the scaling factor $\alpha$ increases for very cool stars with effective temperatures below 4000\,K, which we interpret as spots located at higher latitudes.}
{We conclude that new techniques (such as the GPS method) must be applied to detect rotation periods of stars with small and more irregular variabilities. Ignoring these stars will distort the overall picture of stellar activity and, in particular, solar-stellar comparison studies.}

\maketitle

\section{Introduction}

The stellar rotation period $\Prot$ is a fundamental quantity in stellar astrophysics because it is closely linked to the star's activity level and its age. \citet{Skumanich1972} first demonstrated that the average equatorial rotational velocity and the emission luminosity in the cores of the \ion{Ca}{ii} H and K lines both decrease with stellar age $t$ according $\Prot \sim t^{1/2}$. In the following years it has been shown that, on average, young stars rotate faster and are more active, whereas old stars rotate more slowly and are less active (e.g., \citealp{Noyes1984}). The pioneering work of \citet{Skumanich1972} has inspired the idea of age-dating the star using its rotation period. This semi-empirical method, nowadays known as gyrochronology \citep{Barnes2003,Barnes2007}, calibrates the relation between the stellar mass, rotation period, and age. Hence, knowing the stellar rotation period is essential for estimating the stellar age -- a fundamental quantity of the star that cannot be measured directly.

Most commonly, stellar rotation periods are measured by observing stellar brightness variations over time, and searching for repeatable patterns in long-term (photometric) time series caused by star spots rotating in and out of view. Owing to the \textit{Kepler} space telescope's almost uninterrupted photometric observations of $\sim150,000$ main-sequence stars for 4 years, rotation periods have been measured for several tens of thousands of stars \citep{McQuillan2013a, McQuillan2013b, Reinhold2013, Walkowicz2013, Nielsen2013, McQuillan2014, doNascimento2014, Garcia2014, Reinhold2015, Ceillier2016, Ceillier2017, Santos2019, Santos2021}.

Among these studies, one of the largest samples of rotation periods was provided by \citet{McQuillan2014} (hereafter \citetalias{McQuillan2014}), measuring periodic brightness variations in 34,030 \textit{Kepler} stars, which remains one of the largest collections of rotation periods today, which has been used in numerous studies covering a wide range of topics from constraining stellar dynamo theories to understanding the evolution of our Galaxy (see, e.g., \citealp{vanSaders2019} for one of the most recent examples). Despite this huge number, \citetalias{McQuillan2014} could not unambiguously detect periods in an even larger sample of 99,000 stars. 

Recently, \citet{Santos2019,Santos2021} (hereafter \citetalias{Santos2021}) reanalyzed the full \textit{Kepler} archive and significantly increased the number of detected rotation periods to 55,232 out of 159,442 targets. Both studies \citepalias{McQuillan2014,Santos2021} identified a decrease of the period detection rate with increasing effective temperature. M dwarfs have detection fractions of 70--80\% and K dwarfs around 50\%, whereas the detection fraction drops to $\sim30\%$ or less for F and G dwarfs. This observation can be explained by the fact that the variability pattern changes from cooler to hotter stars. Light curves of M dwarfs show very regular periodicity over many rotation periods, whereas G-type stars exhibit more irregular variability, hardly showing any periodicity over long sections of the observations. It was suggested that the cause for this behavior is that the spot lifetimes of these stars are often shorter than the stellar rotation period, which leads to irregularities in the light curves \citep{Giles2017,Basri2022}.

As a consequence, the rotation periods of many stars around solar spectral type remain undetected in an automated period search because the rotational periodicity is not stable enough to generate a significant peak in the frequency analysis \citep{Reinhold2021}. This implies that the conclusions drawn from many studies of near-solar rotators (e.g. \citealp{vanSaders2019,Reinhold2020,Okamoto2021}) might be strongly biased toward the behavior of more active stars, for which rotation periods could be determined. In particular, the relatively small number of stars with known rotation periods and variabilities similar to that of the Sun conveys a false picture of the Sun being unusually quiet compared to other stars with detected rotation period \citep{Reinhold2020}. 

Our main goal of this study is to make use of recent developments in understanding stellar brightness variability and utilizing new methods to determine rotation periods of a larger sample of \textit{Kepler} stars than ever before. This extended sample of stars with determined rotation periods should at least partly remove these biases. Possible applications range from the comparison of observed period distributions in the \textit{Kepler} field to predictions of Galactic evolution models \citep{vanSaders2019}, comparing solar and stellar variabilities \citep{Reinhold2020}, and the search for superflares on solar-like stars \citep{Okamoto2021,Vasilyev2022}.

To achieve this goal, more rotation period measurements of stars with small, solar-like variabilities are needed. Recently, \citet{GPS_I} showed that the correct rotation period of stars with irregular variability can reliably be detected by a novel method that considers the Gradient of the (global wavelet) Power Spectrum (GPS), instead of the power spectrum itself.

The GPS method has been successfully applied to measure the solar rotation period \citep{GPS_II}. Furthermore, it shows good agreement with the previously reported periods of \textit{Kepler} stars \citep{GPS_III}. We emphasize that, in contrast to classical period analysis methods, the GPS method does not require a repeatable spot pattern in the time series but is sensitive to the typical dip durations of spots crossings (s. Sect.~\ref{methods}). Consequently, it even works in cases when the magnetic features live shorter than the stellar rotation period such that no reoccurring transits of the same magnetic features are required. Hence, the GPS method is ideally suited to measure rotation periods for stars where classical methods failed to detected reliable periods before \citep{Reinhold2022}. 

\section{Data and sample selection}

\subsection{\textit{Kepler} data}
In this work, we analyze the long-cadence light curves processed by the latest version of the \textit{Kepler} pipeline (Data Release~25). The data\footnote{The data can be retrieved at \url{https://archive.stsci.edu/Kepler/data_search/search.php}.} are released in \textit{quarters} with lengths of $\sim90$ days, with exceptions for the quarters Q0, Q1, and Q17 that have shorter observing times between $10-33$ days. In the following, we use all available quarters except for Q0, Q1, and Q17 because these are significantly shorter than the other quarters, which becomes important in the period analysis.

\textit{Kepler} data are known to suffer from various instrumental effects acting on different time scales, and affecting each observing quarter differently strongly. One of the most severe effects are drifts of stars across the detector, leaving long-term up- and downward trends in the light curves. These long-term signals can mimic the variability of slow rotators, and must be treated with caution. Previous attempts cleaned the data from instrumental signals by searching for shared signals across the detector. These so-called cotrending basis vectors are removed from the data by subtracting a linear combination of them from the time series \citep{Kinemuchi2012,Stumpe2012,Smith2012}. It was found that this approach bears the risk of underfitting because instrumental signals were not fully removed. An updated version of the \textit{Kepler} pipeline separates the instrumental systematics by frequency \citep{Stumpe2014}. This approach, however, at times overcorrects the data, and removes true astrophysical signals.

Even though the data used here were reduced with the latest pipeline, visual inspection showed that the reduction was far from being perfect. Many quarters still contained instrumental trends which showed an increased variability compared to the other quarters for a given star. Such instrumental trends are often found every 4th quarter because the \textit{Kepler} telescope rolls by 90 degrees every quarter such that a certain target falls on the same CCD every 4 quarters. To identify such cases in an automated way, we use a common metric that characterizes the light curve variability: the variability range $\Rvar$ \citep{Basri2010,Basri2011}. This measure computes the difference between the 95th and 5th percentile of the sorted differential intensities. Here, we compute the variability range from the 3-hours binned time series for each quarter individually, which we denote by $R_{\rm var,\,q}$, and also compute the median of all quarters $R_{\rm var,\,med}$. After trying different thresholds, we found that all quarters with variabilities $R_{\rm var,\,q} > 3\cdot R_{\rm var,\,med}$ should be discarded from the analysis. A table with all removed quarters can be found in the online version of the paper.

\subsection{Sample}
The GPS method was originally developed and calibrated to measure periods of stars with near-solar effective temperatures, including the solar rotation period (s. Sect.~\ref{methods}). However, it was found that the method also yields reliable periods for stars of later spectral type \citep{GPS_III}. Thus, we use the revised stellar properties catalog of \citet{Mathur2017} and select main-sequence stars with effective temperatures $\Teff \leq 6500$\,K and surface gravities $\logg \geq 4.0$. We further discard all stars matching the \textit{Kepler} eclipsing binary catalog\footnote{The catalog can be found at \url{http://Keplerebs.villanova.edu/}} by \citet{Kirk2016}, as well as 9 stars with residual instrumental systematics (with KIC numbers 6063291, 6126271, 7627042, 7800157, 11393439, 11414728, 11515679, 11805150, 11808713). This selection leaves \nsel stars in total. These rather loose criteria should ensure that the targets lie on the main sequence (or close to it). 

Recently, \citealt{Berger2020} (hereafter \citetalias{Berger2020}) published an updated catalog of fundamental parameters of $\sim 186,000$ \textit{Kepler} stars taking into account Gaia DR2 parallaxes. The fundamental parameters of both catalogs clearly show some deviation. In particular, the \citetalias{Berger2020} temperatures are roughly 200\,K cooler and the \citetalias{Berger2020} surface gravities are on average 0.2 dex smaller than the values given in \citet{Mathur2017}. We find that 113,867 of the selected \nsel stars fulfill the chosen criteria for the parameters given in the catalog of \citetalias{Berger2020}. We note that the main goal of this study is to measure rotation periods for as many stars as possible, and not to decide on the accuracy of the fundamental parameters, which is beyond the scope of this study.

\section{Methods}\label{methods}
\subsection{The GPS method}
First, we prepare the final light curves used in the following analysis. The flux of each observing quarter Q2 to Q16 is divided by its median, subtracted by unity, and outliers are removed which exceed 6 times the median absolute deviation. The time series are appended and the resulting light curve is binned to 3 hours, forming the final light curves used in this analysis. The binning reduces the (photon + granulation) noise in the light curve and 3--6 hours is a typical granulation timescale, dominating the variability in this time interval. We note that rotational variability (i.e., our main focus) starts to dominate on 6 hours and longer time scales.

\begin{figure*}
    \centering
    \includegraphics[width=\textwidth]{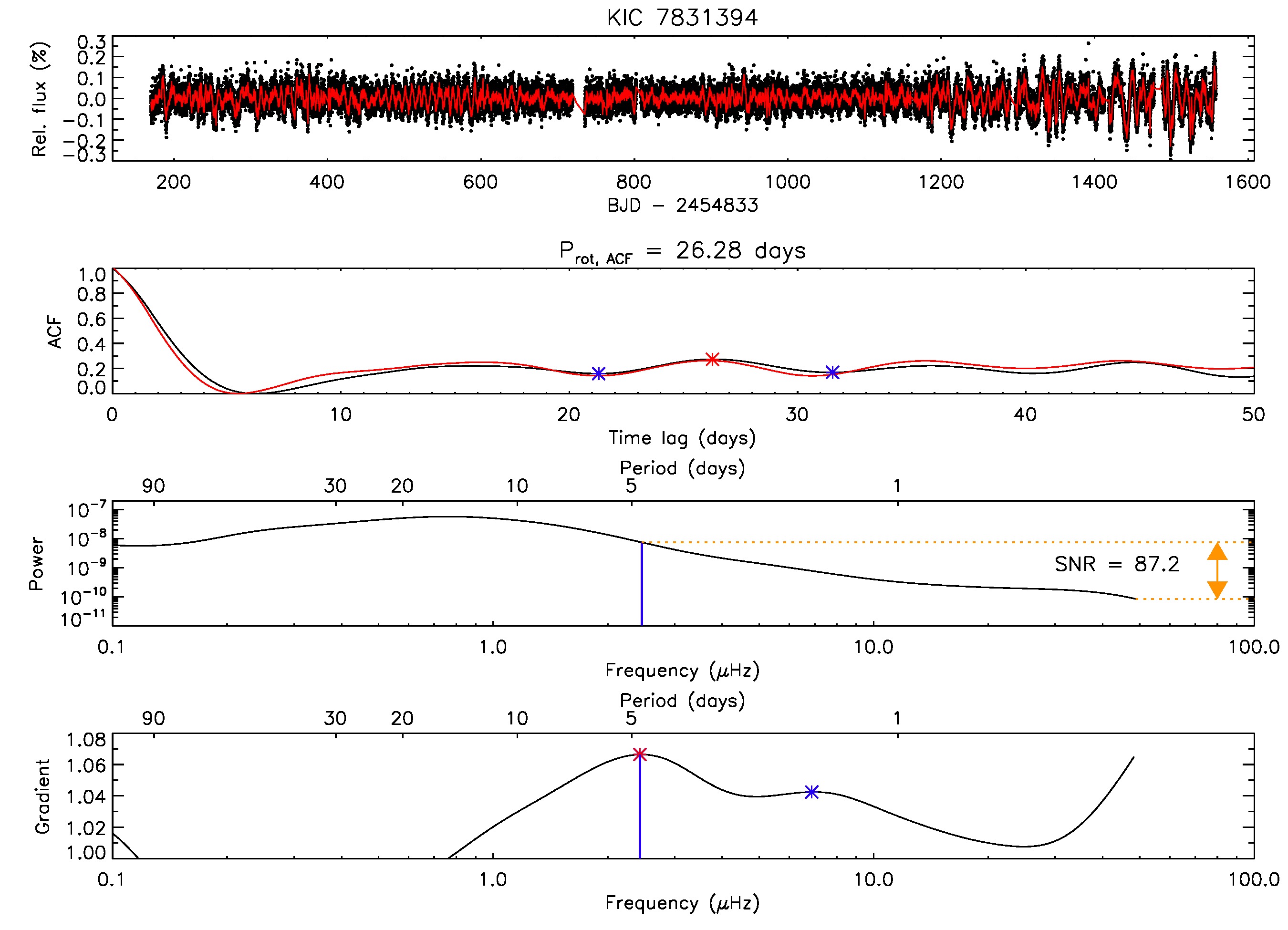}
    \caption{Example of a variable \textit{Kepler} star and different period analysis methods applied to the data. Top panel: Original (black dots) and 3h-binned (red line) light curve of the star KIC\,7831394. Second panel: global (black) and local (red) auto-correlation functions (ACFs). The best ACF period is found at 26.28~days. Third panel: Global wavelet power spectrum (black). The vertical blue line indicates the position of the inflection point. The ratio between the power at the inflection point and the minimum is defined as signal-to-noise ratio (SNR), indicated by the orange arrow between the dotted lines. Bottom panel: Gradient of the Power Spectrum (GPS). The maximum is found at the inflection period $\Pip=4.76$ days, marked by the blue line and the red asterisk.}
    \label{KIC7831394}
\end{figure*}

An example light curve of the star KIC\,7831394 is shown in the top panel of Fig.~\ref{KIC7831394}. The light curve clearly shows variability on rotational time scales (especially after binning over 3 hours), making it a promising candidate for our period analysis. In the second panel of Fig.~\ref{KIC7831394}, we compute the auto-correlation function (ACF) using the IDL function A\_CORRELATE. We further subtract the ACF minimum and normalize it by dividing out the maximum such that all ACF values lie between zero and one. For stars with periodic variability patterns, the ACF has proven to be a good tool for measuring the rotation period. The ACF of the full time series is shown in black. The highest peak is found at a period of $P_{\rm rot,\,ACF}=26.28$ days. As goodness measure of the period, we compute the local peak height (LPH) as the difference between the highest peak (red asterisk) and the mean of the two troughs on either side (see, e.g., \citealt{Reinhold2021} for details).

Additionally, we compute the ACF for each quarter Q2 to Q16 individually and compute the mean ACF power in 0.1 day bins. This function is referred to as the \textit{local} ACF, in contrast to the \textit{global} ACF described above. The local ACF is shown in red on top of the global ACF in Fig.~\ref{KIC7831394}. For quite periodic light curves, both functions are very similar. However, for stars with more irregular variability, the local and global ACF often differ. As a consequence, the highest peak of the one does not match the one of the other, and the period is unclear. We return to this point in Sect.~\ref{goodness}.

Now we compute the wavelet power spectrum of the full time series (third panel in Fig.~\ref{KIC7831394}). The highest power is found at a period of $\sim15.17$ days. As we will see later, this period is likely an artifact of the high-pass filtering of the \textit{Kepler} data. The bottom panel of Fig.~\ref{KIC7831394} shows the gradient of the power spectrum (GPS). The gradient is computed using eq.~3 in \citealt{GPS_I} (for details, we refer the reader to \citealt{GPS_I}). The maximum of the gradient corresponds to the position of the inflection point (IP), i.e., the point where the curvature of the high-frequency tail of the power spectrum changes its sign. The period at this inflection point, $\Pip$, is linked to the stellar rotation period $\Prot$ by the simple equation
\begin{equation}\label{eq1}
    \Prot = \Pip / \alpha
\end{equation}
where $\alpha$ is a calibration factor (s. Sect.~\ref{results}). The main idea behind the GPS method is that the high-frequency tail of the power spectrum is much less affected by the evolution of magnetic features than the power spectrum peak associated with the rotation period (see Fig.~3 in \citealp{GPS_I}). In this case, the inflection period is found at $\Pip=4.76$ days, indicated by the blue line and the red asterisk. This period is linked to the rotation period by eq.~\ref{eq1}. Using the calibration factor $\alpha=0.217$ derived by \citet{Reinhold2022} yields a rotation period $P_{\rm rot,\,GPS}=21.94$ days. 

This result is in good agreement with the rotation period of 22.11 days derived by \citetalias{Santos2021}. However, the period $P_{\rm rot,\,GPS}$ shows some discrepancy to the ACF period, and is very different from the one at the highest peak of the power spectrum. As we will see later, the ACF method often fails to detect the correct rotation period, the more irregular the variability gets.

\subsection{Goodness of the GPS periods}\label{goodness}
For each light curve, the GPS method returns an inflection point period. However, it is not always clear if this period can be associated with the rotation of active features over the stellar surface. Thus, we define different goodness metrics for the derived GPS and ACF periods. Eventually, these metrics are combined to a point system that assigns a certain number of points to each star to assess the period reliability: the higher the number of points, the more periodic the signal. 

Most commonly, classical period analysis methods define the highest peak of the power spectrum (or the first ACF peak) as the strongest periodicity in the data. It was found by visual inspection of many different \textit{Kepler} stars that the highest GPS peak nicely scales with the periodicity in the light curve. If the highest peak lies in the range 0.5--10 days, we save the inflection point period $\Pip$ and the associated peak height $\hip$. The lower limit of 0.5 days is basically determined by the 3h binning of the data (the Nyquist frequency would be 1/6h), whereas the upper limit of 10 days should prevent running into problems with the data reduction (see Fig.~\ref{Prot_ACF_GPS_dist} and subsequent discussion). According to Eq.~\ref{eq1}, the considered range of inflection periods enables us to detect rotation periods in the range $\approx 2.3-46.7$ days. 

To compute the power spectra, we use the IDL function \textit{WV\_CWT}, which returns the continuous wavelet transform, and set the keyword dscale=1/32, which affects the peak height values. In this normalization, the peak height distribution $\hip$ ranges from $1-1.15$, with a median height of $1.05$. Strictly periodic stars have large peak heights $\hip>1.06$ to which we assign 1 point. Less periodic but still variable stars have peak heights $1.04<\hip<1.06$ to which we assign 0.5 points. Light curves that are completely dominated by noise have even smaller peak heights and, thus, get 0 points. 

Even though it was shown that the highest power spectrum peak itself is not necessarily a good measure of the rotation period, we can still use the power spectrum to define another goodness measure: we call the ratio between the power at the inflection point and the minimum power of the spectrum the signal-to-noise ratio SNR (see third panel in Fig.~\ref{KIC7831394}). Similarly to $\hip$, this quantity also scales with the periodicity. By visual inspection of many light curves and the SNR distribution of all stars, we assign 1 point if $\rm SNR>50$, 0.5 points if $10<\rm SNR<50$, and 0 points otherwise.

For the ACF, we have already defined the LPH as a goodness metric (see, e.g., \citealt{Reinhold2021}). These authors found that strong periodicity is usually found for $\rm LPH>0.2$ (1 point). Less periodic time series still reach values $0.1<\rm LPH<0.2$ (0.5 points), and purely noisy stars exhibit small $\rm LPH<0.1$ (0 points). We note that the LPH used here always refers to the global ACF. Additionally, we compare the global and the local ACF period. If these two periods agree within 10\%, we add another 0.5 points.

\begin{figure*}
  \centering
  \includegraphics[width=0.47\textwidth]{
  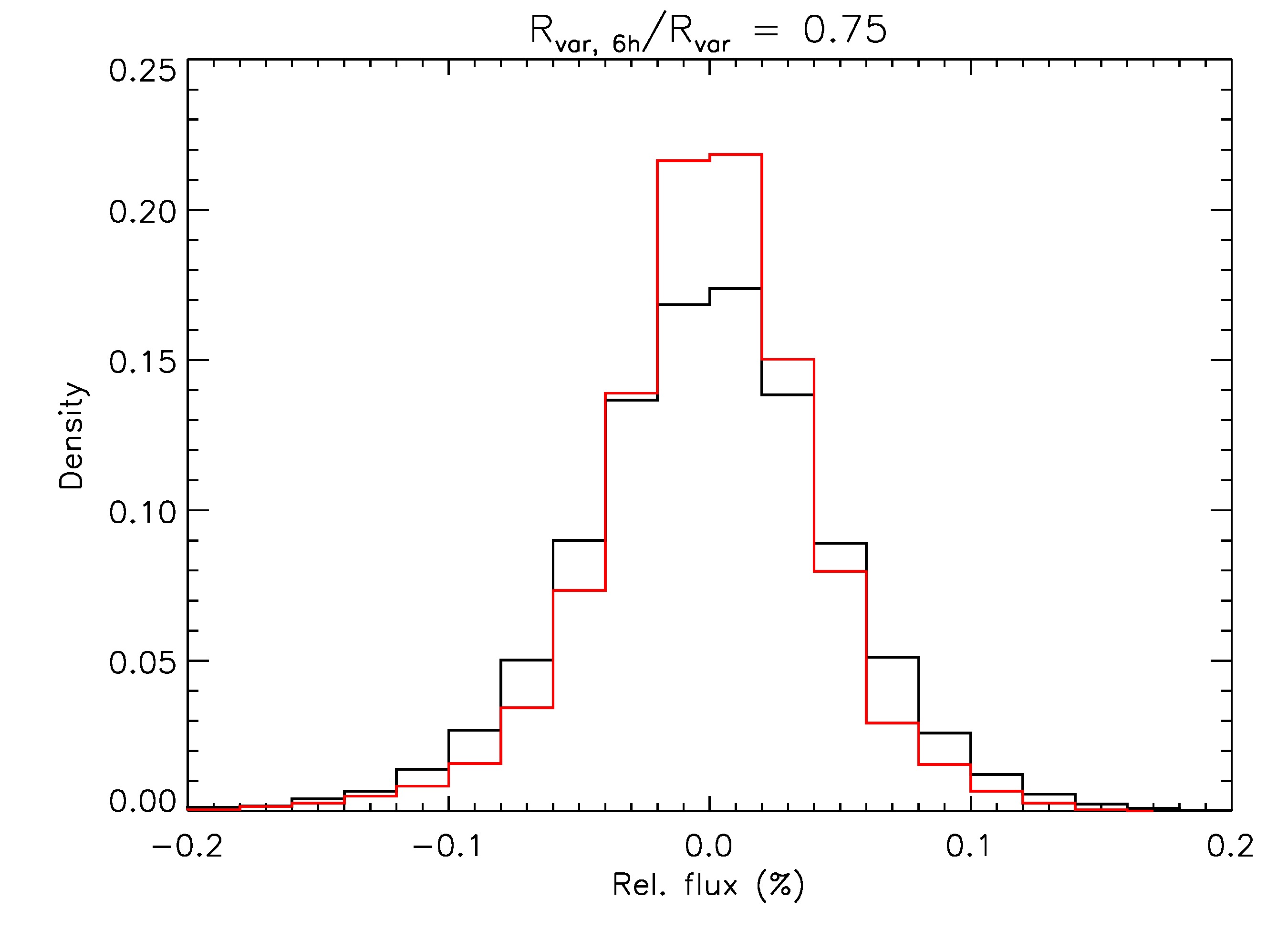}
  \includegraphics[width=0.47\textwidth]{
  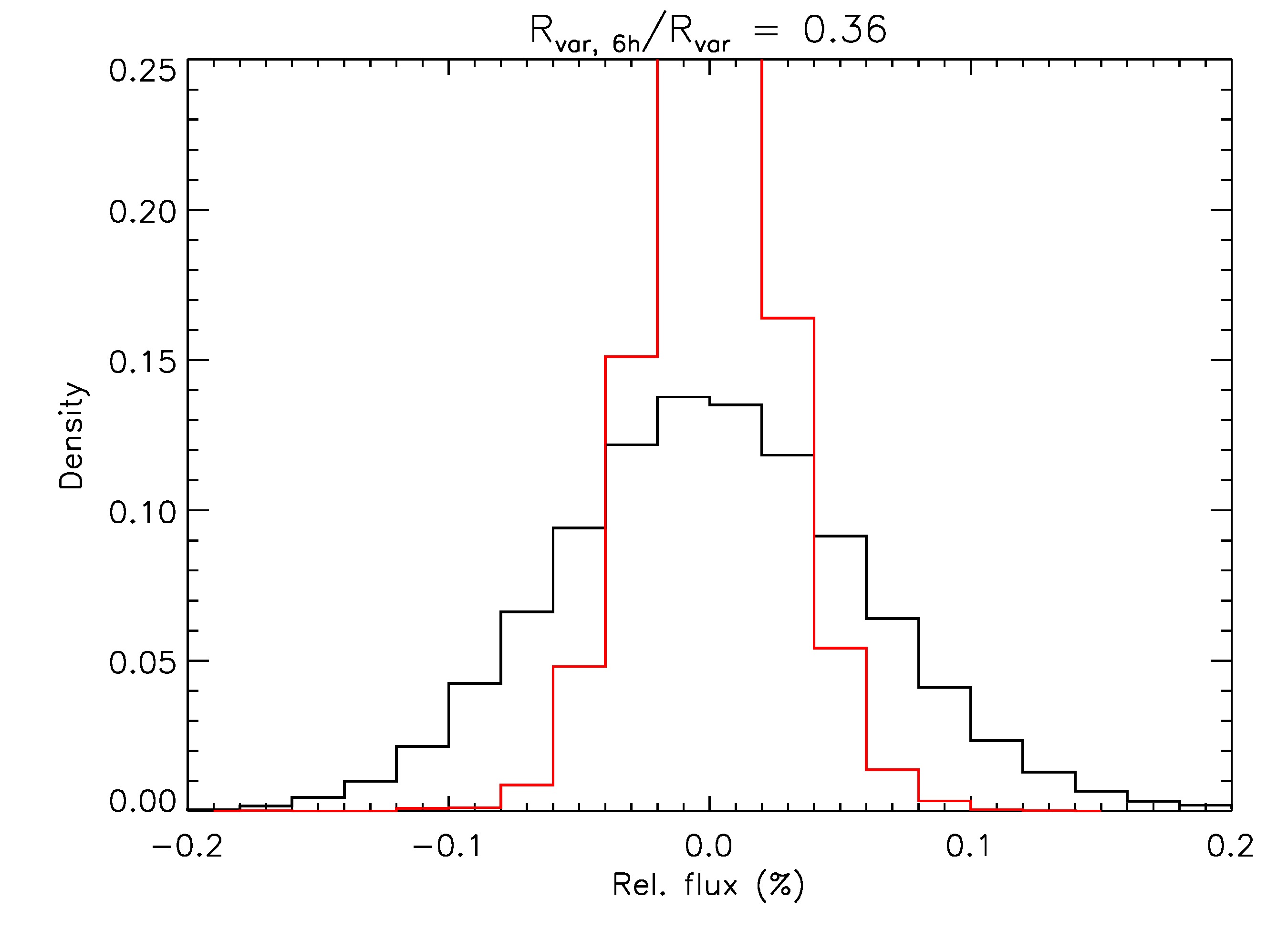}
  \caption{Distribution of the 6h-binned (red) and unbinned (black) flux values of the stars KIC\,7831394 (left) and KIC\,11802969 (right).}
  \label{Rvar_frac}
\end{figure*}

By visual inspection it was found that many light curves are dominated by noise and hardly show any variability, even less periodicity. Nevertheless, all metrics above will return some values since they respond to any signal (even to pure noise) in the time series. However, these very quiet stars can be identified by comparing the variabilities of the unbinned and binned time series. In Fig.~\ref{Rvar_frac}, we show the relative fluxes of the unbinned (black) and 6-hours binned (red) data as a histogram. The left panel of Fig.~\ref{Rvar_frac} shows that both flux histograms nicely overlap (also see top panel in Fig.~\ref{KIC7831394}), which means that the light curve of the star KIC\,7831394 is dominated by (rotational) variability. On the contrary, the binned and unbinned flux distributions of the star KIC\,11802969 (right panel) look very different. The 6-hours binning reduced the noise in the light curve and the remaining variability is small, which means that the photometric variability of this star is completely dominated by noise. 

Instead of looking at flux distributions, we can simply compute the variability ranges $\Rvar$ of the unbinned and 6-hours binned time series, and compute their ratio. If the $R_{var,\,6h}/R_{var}$ ratio is close to unity, the light curve variability is dominated by rotation. By visual inspection, we assign 1 point if $R_{var,\,6h}/R_{var}>0.6$, 0.5 points if $0.4<R_{var,\,6h}/R_{var}<0.6$, and 0 points otherwise. We note that also the 3-hours binned value $R_{var,\,3h}$ could have been used instead but the 6h-binning reduced the noise even stronger.

Combining all metrics defined above yields points in the range 0 (pure noise) to 4.5 (highest periodicity) for each star. For the sake of clarity, the point metric is summarized in Table~\ref{points_table}. We will return to this point system in Sect.~\ref{results}. 

\begin{table}
\begin{tabular}{cccc}
\hline
points & 0 & 0.5 & 1 \\
\hline
$\hip$ & $<1.04$ & $1.04-1.06$ & $>1.06$ \\
SNR & $<10$ & $10-50$ & $>50$ \\
LPH & $<0.1$ & $0.1-0.2$ & $>0.2$ \\
$R_{var,\,6h}/R_{var}$ & $<0.4$ & $0.4-0.6$ & $>0.6$ \\
\hline
\end{tabular}
\caption{Point system for the individual goodness metrics.}
\label{points_table}
\end{table}

\section{Results}\label{results}
\begin{figure*}
    \centering
    \includegraphics[width=0.85\textwidth]{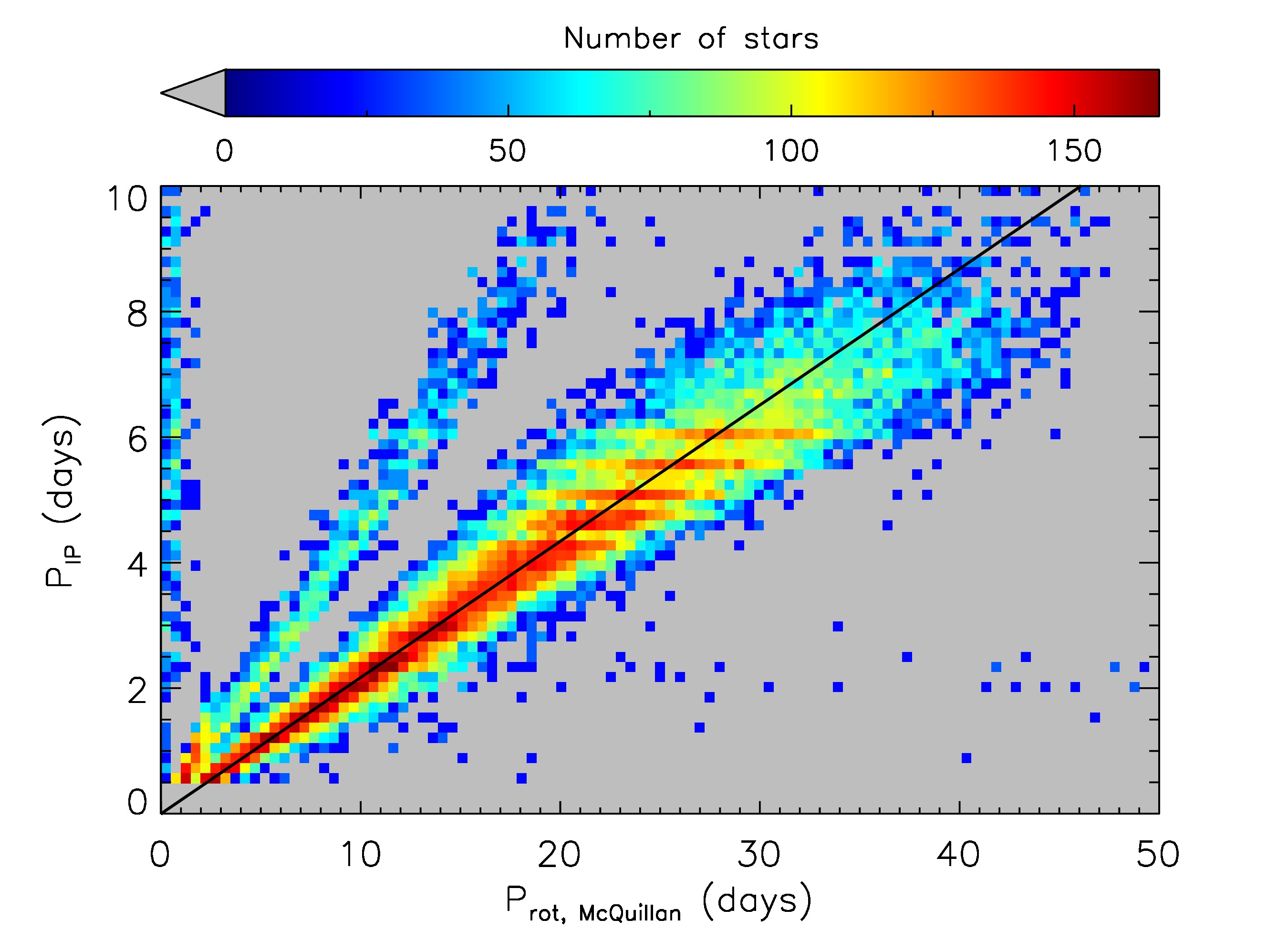}
    \includegraphics[width=0.85\textwidth]{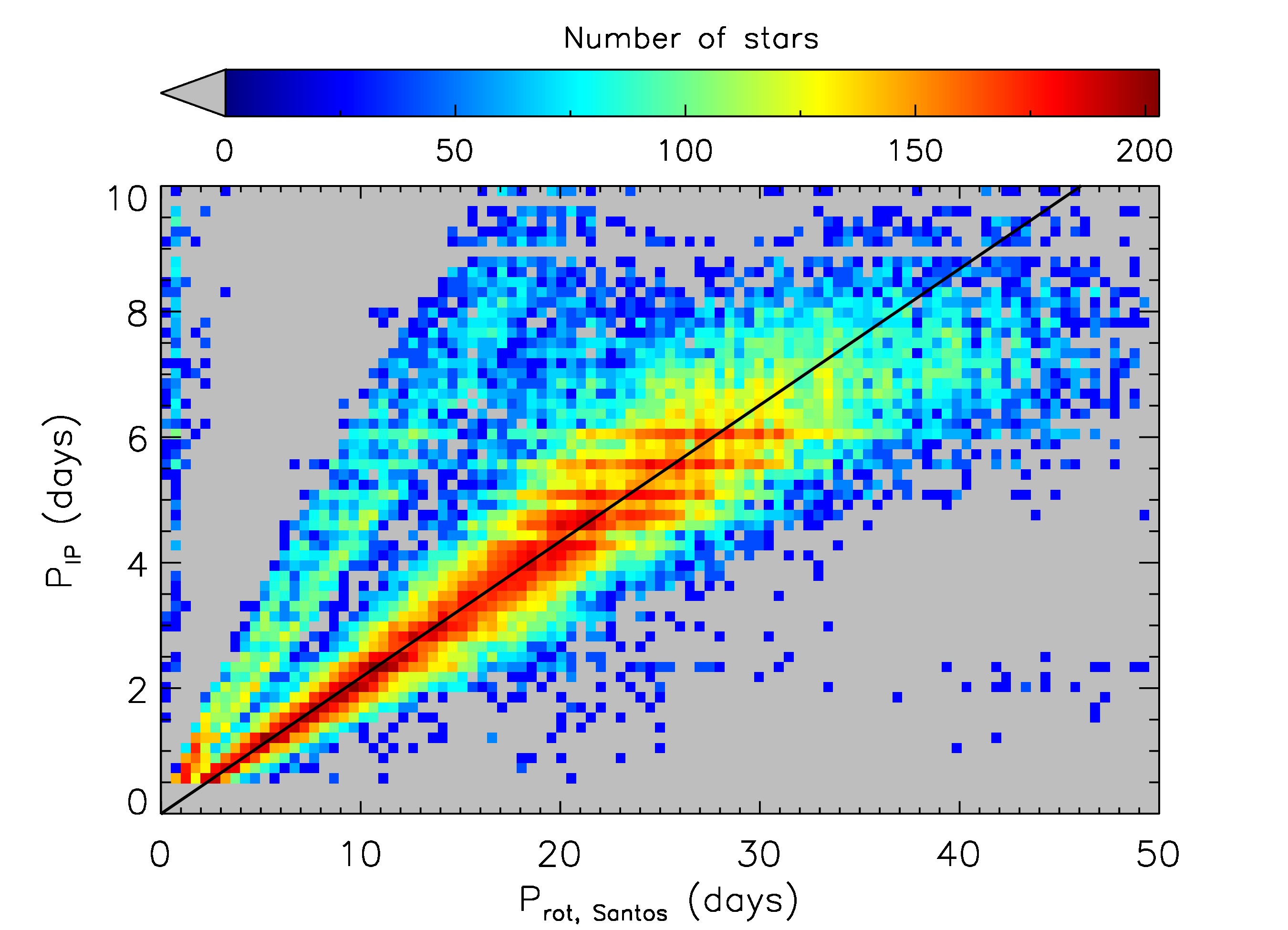}
    \caption{Inflection point period $\Pip$ vs. rotation period derived by \citetalias{McQuillan2014} for 30582 stars (upper panel) and \citetalias{Santos2021} for 48264 stars (lower panel). The solid black line shows the relation given in Eq.~\ref{eq1} with $\alpha=0.217$. Both panels exhibit an upper branch where the GPS method detects (slightly more than) the double of the inflection period.}
    \label{density_plot}
\end{figure*}

For the \nsel stars in our sample, we could measure an inflection point period within $0.5-10$ days for \nseltwo stars. We note that a fraction of this sample was assigned zero points in the end (compare Fig.~\ref{pts_dist}). We now compare our results to a sample of stars with previously determined periods. This comparison will show whether the calibration factor $\alpha=0.217$, previously determined for models of solar-like stars \citep{Reinhold2022}, still holds for real data. In Fig.~\ref{density_plot}, we show the measured inflection periods $\Pip$ against the rotation periods derived by \citetalias{McQuillan2014} (upper panel) and \citetalias{Santos2021} (lower panel) for the stars in common. The solid black line shows the relation given in Eq.~\ref{eq1} with the calibration factor $\alpha=0.217$. Both panels clearly show the linear dependence between both periods for the vast majority of stars. However, a second branch with (slightly more than) twice the inflection period, and consequently twice the rotation period, is visible in both panels. The origin of this \textit{double period branch} is discussed in Sect.~\ref{spot_models}. We further note that the agreement of our period measurements ($\Pip/\alpha$) becomes weaker for $P_{\rm rot,\, Santos}>30$ days. This is caused by the fact that the PDC-MAP pipeline does not preserve variability on these timescales, so the periods retrieved by classical periods analysis tools (as used in \citetalias{McQuillan2014} and \citetalias{Santos2021}) are less reliable (also see Sect.~\ref{Sect_ACF_GPS}).

\begin{figure*}
  \centering
  \includegraphics[width=0.47\textwidth]{
  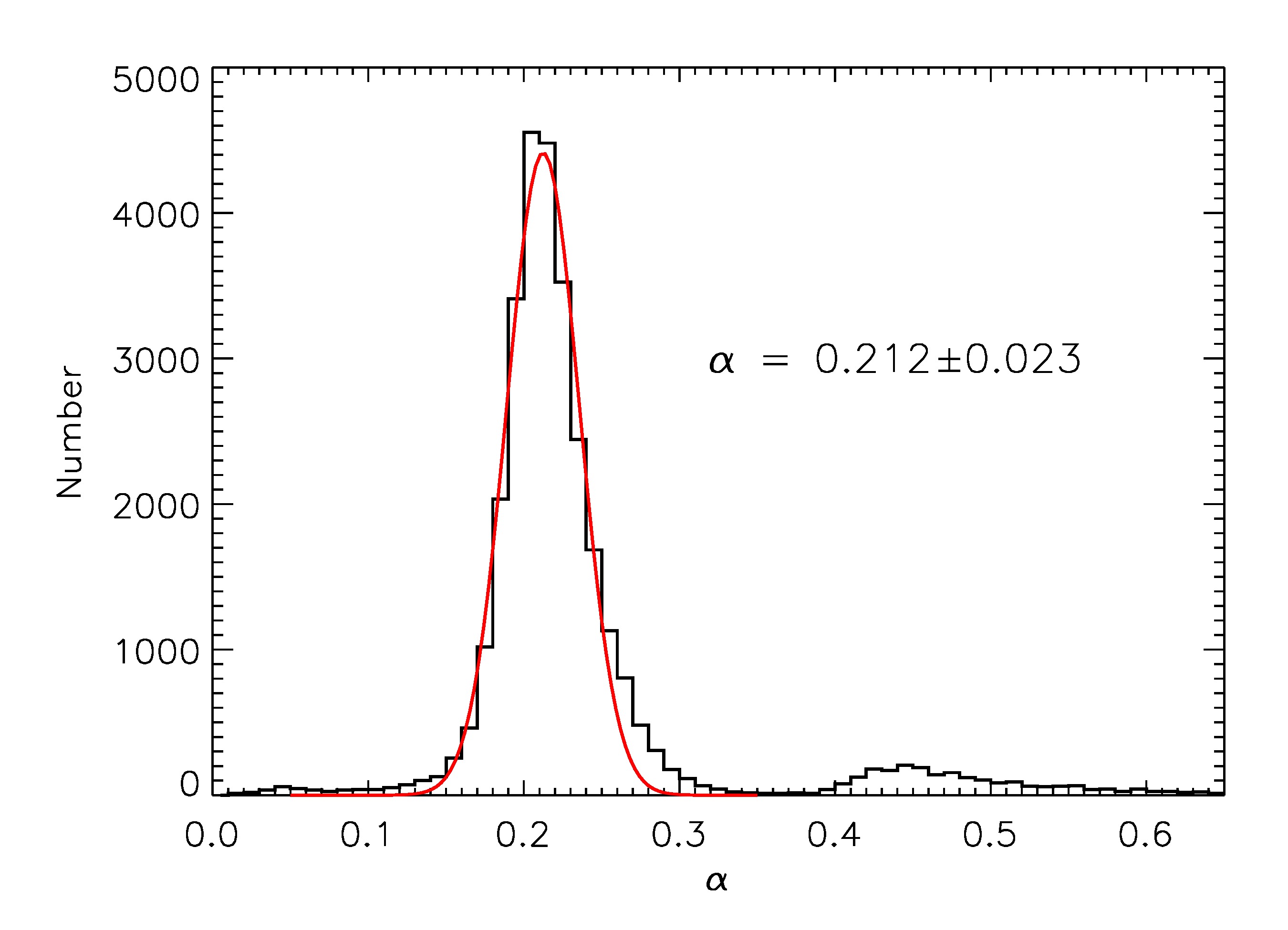}
  \includegraphics[width=0.47\textwidth]{
  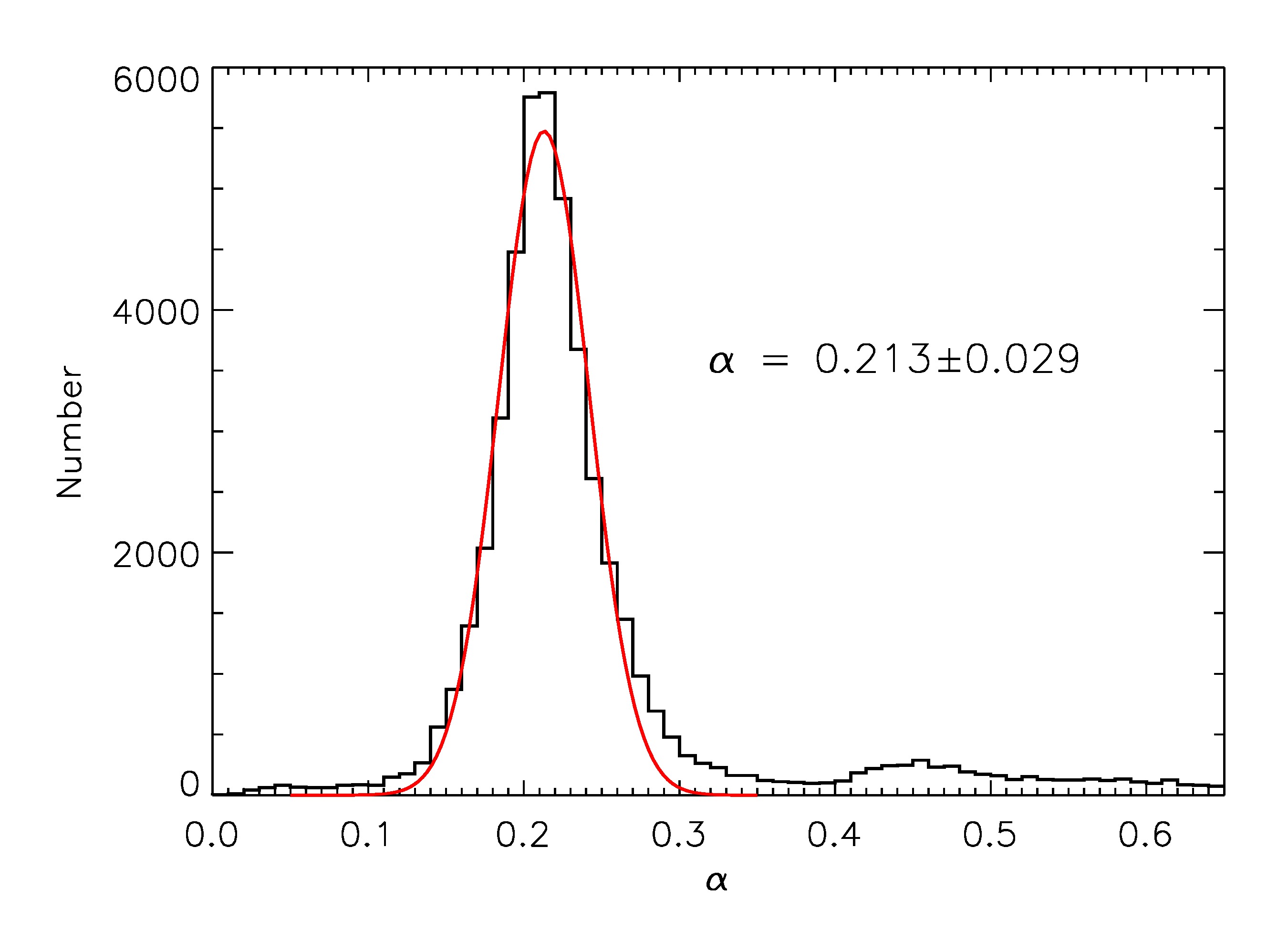}
  \caption{Distribution of $\alpha=\Pip/\Prot$ for the same stars as shown in Fig.~\ref{density_plot} for the rotation periods $\Prot$ of \citetalias{McQuillan2014} (left panel) and \citetalias{Santos2021} (right panel). The red curve shows a Gaussian fit with mean $\langle\alpha\rangle=0.212$ and standard deviation $\sigma_\alpha=0.023$ (left panel) and $\langle\alpha\rangle=0.213$ and $\sigma_\alpha=0.029$ (right panel).}
  \label{alpha_dist}
\end{figure*}

\begin{figure}
  \resizebox{\hsize}{!}{\includegraphics{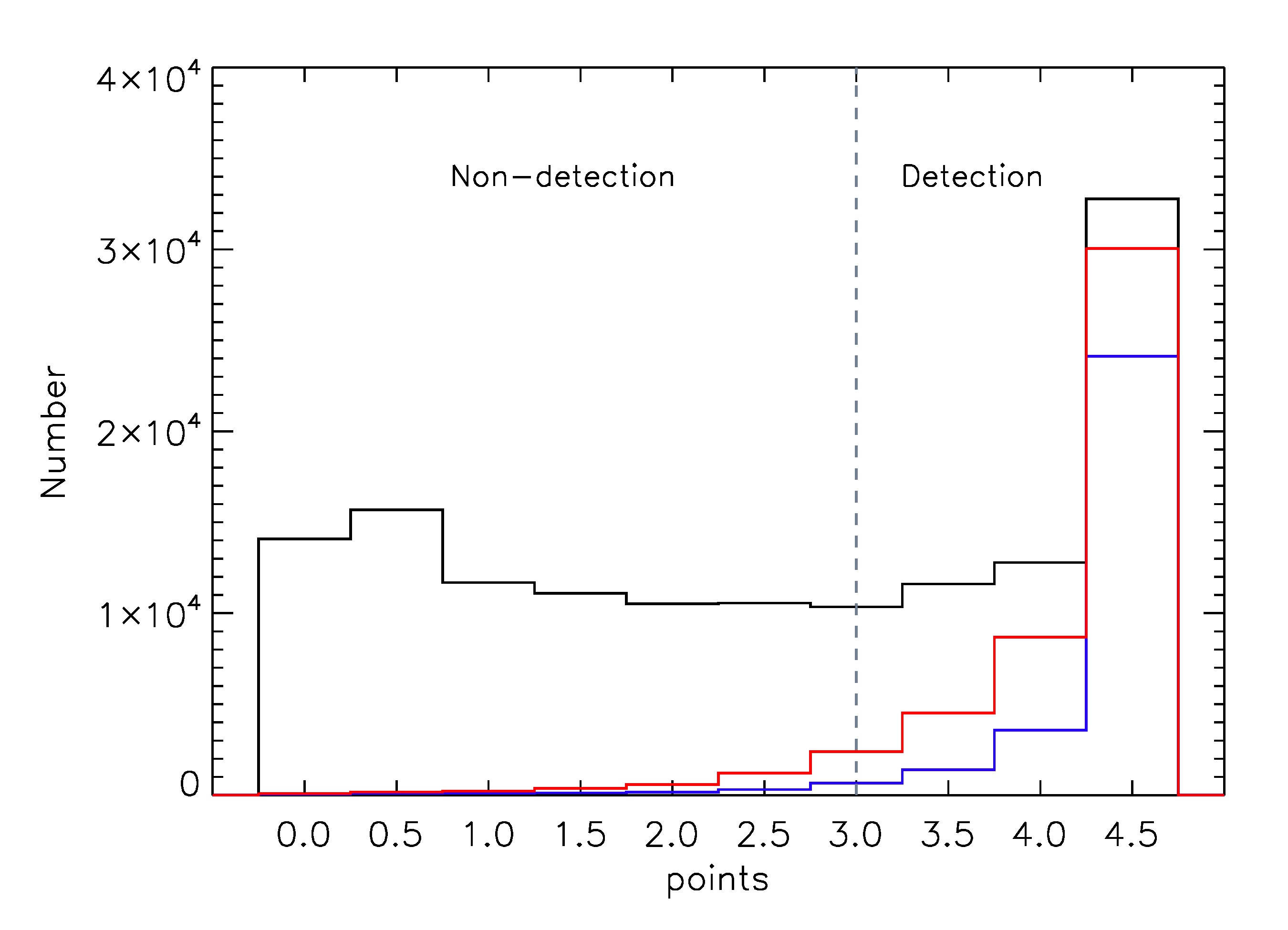}}
  \caption{Distribution of the points metric for all \nseltwo stars with measured inflection period (black), and the ones in common with \citetalias{McQuillan2014} (blue) and \citetalias{Santos2021} (red).}
  \label{pts_dist} 
\end{figure}

Fig.~\ref{density_plot} revealed that there is indeed a linear dependence between the inflection period $\Pip$ and the rotation period $\Prot$. However, this relation is accompanied by large scatter. As mentioned in \citet{GPS_I}, and later shown in detail by \citet{Reinhold2022}, the calibration factor $\alpha$ has an intrinsic uncertainty of $\sim 25\%$. In Fig.~\ref{alpha_dist}, we show the ratio of the inflection period to the rotation period derived by \citetalias{McQuillan2014} (left panel) and \citetalias{Santos2021} (right panel). The distributions have a Gaussian shape centered at $\langle\alpha\rangle=0.212$ (left) and $\langle\alpha\rangle=0.213$ (right) with standard deviations $\sigma_\alpha=0.023$ (left) and $\sigma_\alpha=0.029$ (right). These values are in good agreement with the value $\alpha=0.217$ derived by \citet{Reinhold2022}. The small bump at twice these $\alpha$ values is associated with the upper branch in Fig.~\ref{density_plot}.

We now turn to the point system defined at the end of Sect.~\ref{methods}. In Fig.~\ref{pts_dist} we show the distribution of points allocated to each star. The sample of \citetalias{McQuillan2014} is shown in blue and the stars in common with \citetalias{Santos2021} are shown in red. It is obvious that the number of stars with previously determined periods steeply increases with the number of points, and that the vast majority of those stars has the highest possible number of points. That means that these stars exhibit very periodic light curves where the periodicity is picked up easily with standard tools.

As with almost every frequency analysis tool, setting thresholds for the detected peaks or the period significance is quite subjective. Visual inspection of many different quiet and active stars led us to count all stars with $\geq 3$ points as period detections. This threshold requires that at least one of the metrics has 1 point assigned. It is therefore not very conservative but a reasonable choice. We further note that below this threshold only very few rotation periods have been found in the surveys of \citetalias{McQuillan2014} and \citetalias{Santos2021}, which makes it unlikely that many stars with measurable period have been missed. Although one cannot completely rule out the possibility that there exist stars with measurable periods below 3 points, lowering the threshold would make the periods less reliable. In total, the threshold is satisfied for $\nper$ stars, which we focus on in the following.

\begin{figure}
  \resizebox{\hsize}{!}{\includegraphics{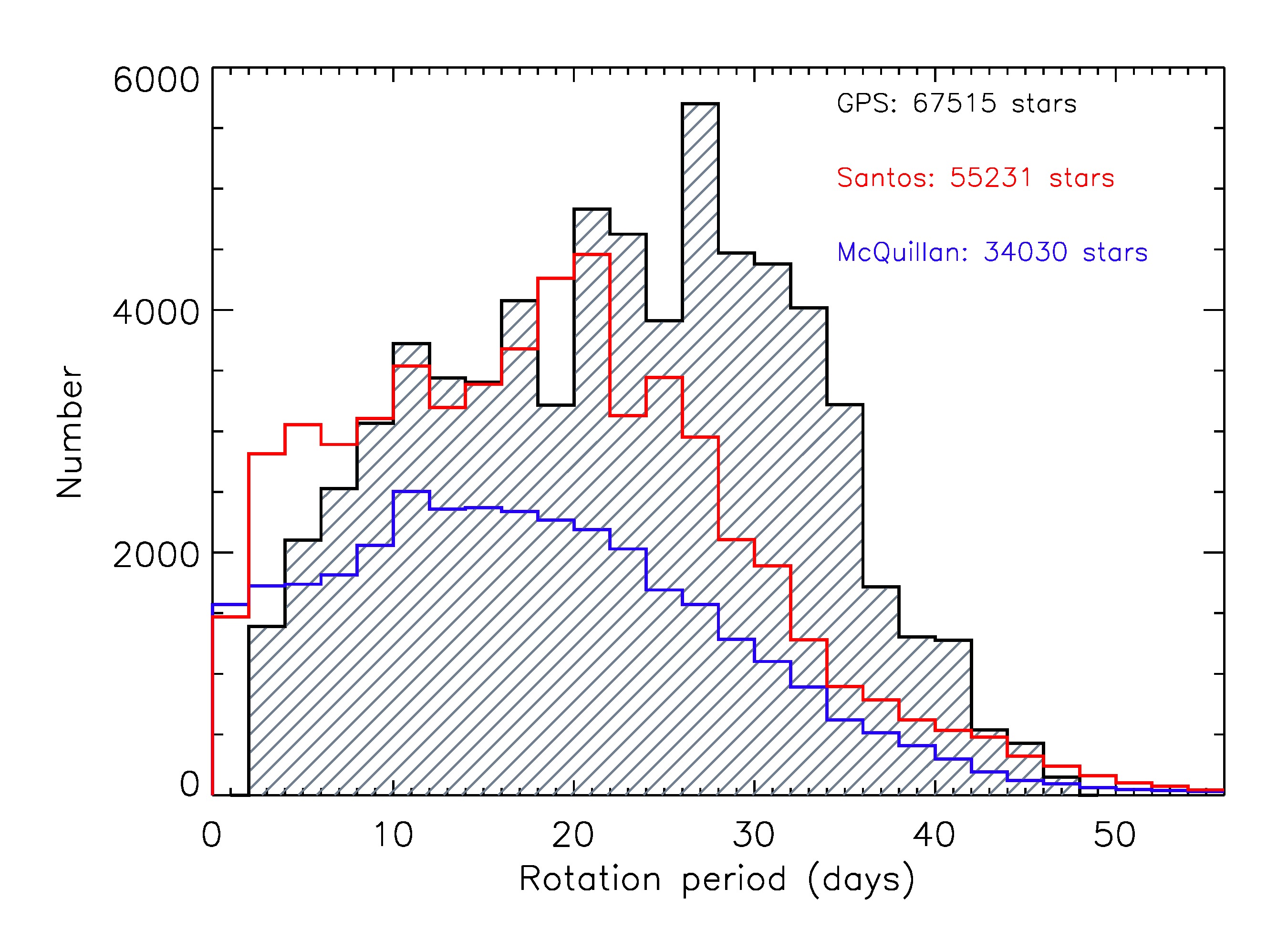}}
  \caption{Rotation period distribution of the \nper stars with points $\geq 3.0$. The rotation periods derived by \citetalias{McQuillan2014} and \citetalias{Santos2021} are shown in blue and red, respectively.}
  \label{Prot_dist} 
\end{figure}

In Fig.~\ref{Prot_dist} we show the distribution of $\nper$ GPS rotation periods of all stars with $\geq 3.0$ points. Here we used $\alpha=0.213$, derived in Fig.~\ref{alpha_dist}, to retrieve the rotation period. The periods determined by \citetalias{McQuillan2014} and \citetalias{Santos2021} are shown as blue and red curves, respectively. Compared to the previous surveys, the GPS method retrieves periods of a larger number of stars. 

In particular, \newper new periods were detected that have not been reported before. Most of the newly determined periods are longer than $\sim28$ days. Furthermore, the median variability of these \newper stars with newly detected periods equals $R_{\rm var,\,3h}=0.085\%$, which is very close to the solar value $R_{\rm var,\,Sun}=0.07\%$ (compare \citealt{Reinhold2020}), and so much smaller than the average variability of all \nper stars ($R_{\rm var, \,3h}=0.17\%$). We emphasize that the detection of these stars with \textit{near-solar} rotation periods and variabilities is a clear benefit of the GPS method.

The reasons why these periods have been missed in previous surveys are manifold: \citetalias{McQuillan2014} used quite conservative thresholds to detect periods, which removes many less periodic slow rotators. Moreover, these authors only considered Q3-Q14 data. Since that time, also the \textit{Kepler} pipeline changed several times. \citetalias{Santos2021} analyzed light curves reduced with their own pipeline as well as those reduced with the latest version (DR25). These authors also combined different period analysis tools and used a machine-learning approach to finally detect periodicity. The GPS method could detect even more (and longer) periods because it does not require a repeatable spot pattern as the other techniques do. The very short periods, however, cannot be accessed because we set a lower limit of 0.5 days to the inflection period, which translates into a lower rotation period limit of $\approx 2.3$ days.

We also checked the cases where a period was reported by 
\citetalias{McQuillan2014} and/or \citetalias{Santos2021} but not detected in this survey. There exist 11,094 periods in these surveys that do not have a period reported here. The majority of these stars were initially not considered in this study because they either had effective temperatures greater than 6500\,K or $\logg<4.0$ dex. Only 3313 out of the 11,094 stars were considered in this study but had a point number smaller than 3.0. Of those, 1064 stars had a reported period outside the considered range of inflection periods between 0.5--10 days, and thus could not have been detected. However, the point distribution of the remaining 2249 stars continuously increases toward our lower limit of 3.0, with a mean number of 2.0 points. This result indicates that the imposed point limit might be lowered, which would eventually lead to even more period detections.

\begin{figure}
  \resizebox{\hsize}{!}{\includegraphics{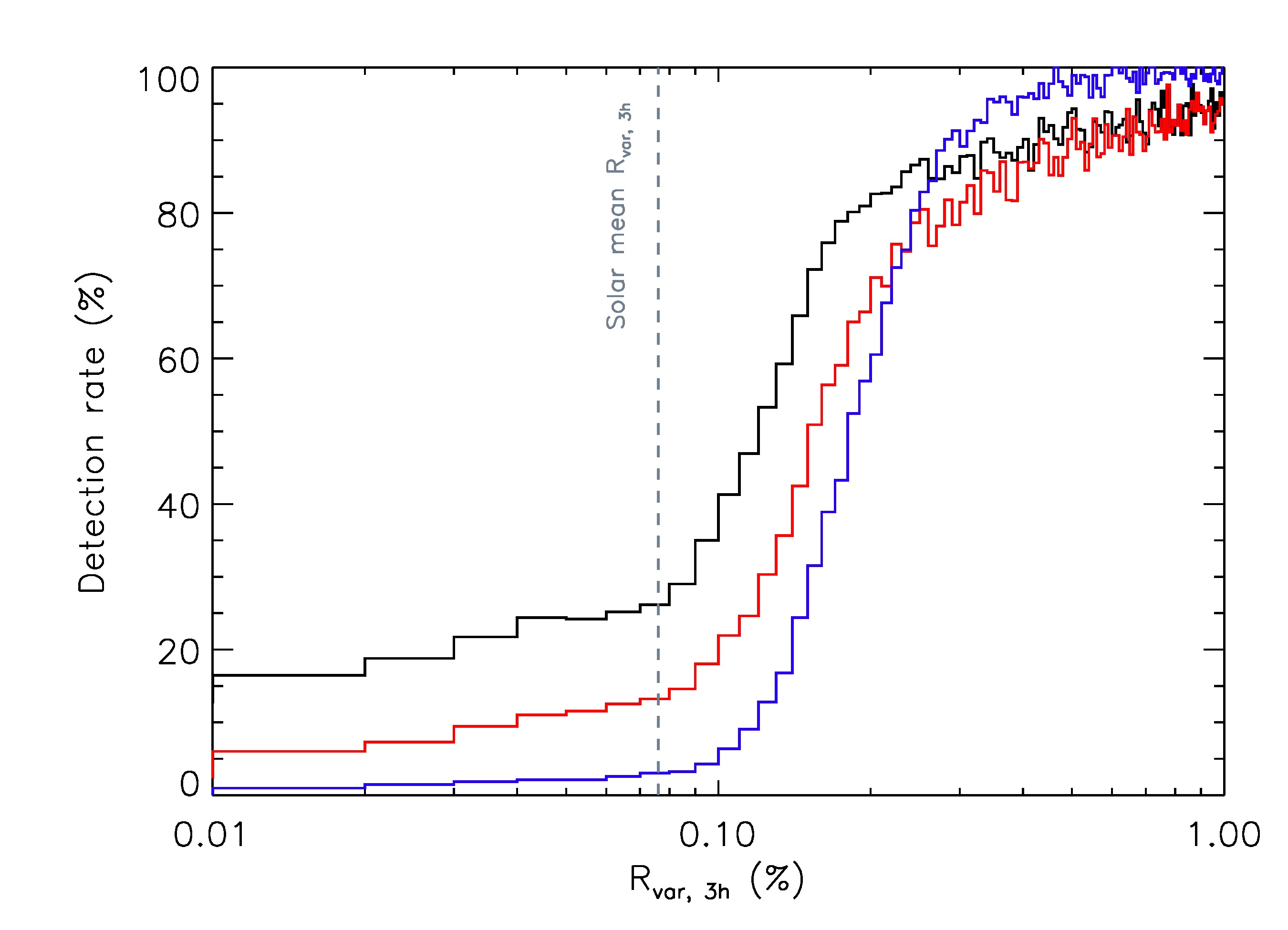}}
  \caption{Detection rate as a function of the variability range $\Rvar$. The black curve shows the detection rate of the GPS method, and the and the detection rates of the methods employed by \citetalias{McQuillan2014} and \citetalias{Santos2021} are shown in blue and red, respectively.}
  \label{detection_rate} 
\end{figure}

The better performance of the GPS method can also be seen in Fig.~\ref{detection_rate}. Here, we show the detection rate as a function of stellar variability. In general, the detection rate increases with variability (without any variability, nothing can be detected). Interestingly, the steepest rise happens shortly after the solar mean variability (gray dashed line). This observation shows that the spot signals - compared to the photon noise in the light curves - start to dominate at this variability. We further see that the black and the red curves show very similar qualitative behavior: both curves steeply increase at small, solar-like variabilities $\Rvar>0.1\%$, and level off at a detection rate of $\approx 93\%$ for variabilities $\Rvar>1\%$. However, it is obvious that the GPS method detects much more periods for stars with smaller variability. 

The detection rate of the \citetalias{McQuillan2014} sample (blue curve) shows a slightly different behavior for variabilities $\Rvar>0.3\%$, where it breaks through the black and red curves, and eventually reaches almost 100\% for the most variable stars. As mentioned above, a different pipeline has been used in \citetalias{McQuillan2014} that better preserved stellar variability. However, this cannot be the reason for the higher detection rate because all values used in Fig.~\ref{detection_rate} have been computed from the latest pipeline used in this study. We attribute the even higher detection rate of the variable stars to an extensive visual light curves inspection of \citetalias{McQuillan2014}, in contrast to the purely automated approach applied by \citetalias{Santos2021} and in this study.

\subsection{GPS vs. ACF periods}\label{Sect_ACF_GPS}
The rotation periods in \citetalias{McQuillan2014}, and to a large extent also those in \citetalias{Santos2021}, have been determined by the auto-correlation function (ACF). Thus, we also computed ACF periods for each star, with an upper period limit of 70 days. In this Section, we compare the ACF and the GPS periods with each other to test the performance of both methods, and to show their limitations.

\begin{figure*}
  \centering
  \includegraphics[width=0.7\textwidth]
  {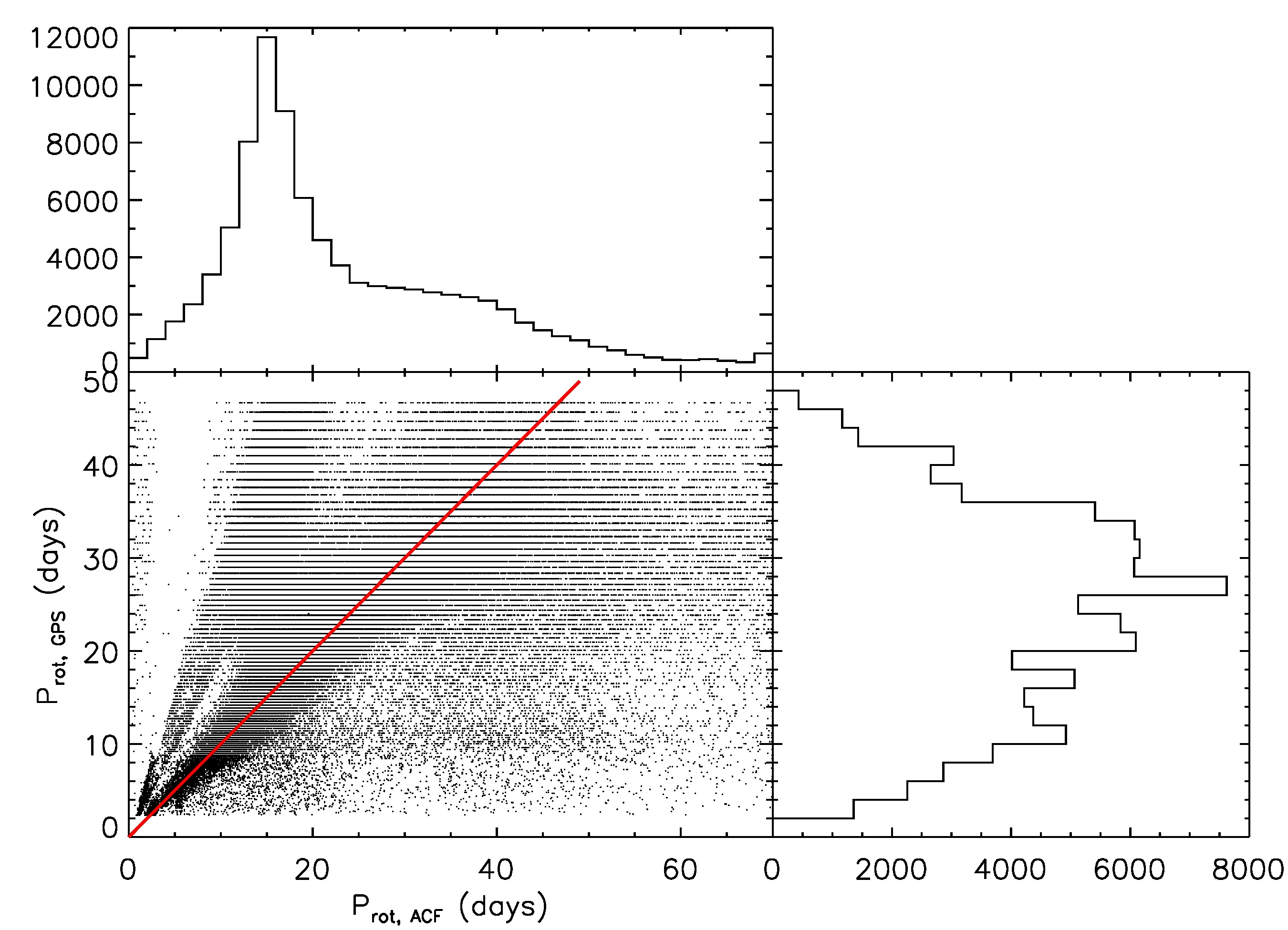}
  \caption{ACF vs. GPS periods for more than 90,000 stars with $\rm LPH>0.1$, with the associated histograms to their sides. The GPS periods have been calculated using the same $\alpha=0.213$ as in Fig.~\ref{Prot_dist}. The red solid line shows the 1:1 identity.}
  \label{Prot_ACF_GPS_dist} 
\end{figure*}

Fig.~\ref{Prot_ACF_GPS_dist} shows the ACF against the GPS periods for all stars where both periods could be measured and the local peak height (LPH) of the ACF peak at least fulfills the mild criterion $\rm LPH>0.1$. For periods less than 20 days there is good overlap between the two methods, as indicated by the 1:1 line (red). Also the secondary branch at twice the ACF period is visible.

A striking feature is certainly the pile-up of ACF periods around 15~days, best visible at the top histogram. We attribute these periods to the high-pass filtering of the data in the latest data reduction, and emphasize that most of these periods are of instrumental origin because such an accumulation of periods is not seen for the GPS periods. It is important to note that there are actually cases where also the GPS returns a period around 15 days, and this periodicity is clearly seen in the light curves. We conclude that the ACF periods in the range 10--20 days cannot be trusted without independent confirmation by another method. The pile-up of periods at around 15 days was also noted and dismissed by \citet{Basri2022}, but without the benefit of this independent analysis. 

\begin{figure*}
  \centering
  \includegraphics[width=0.9\textwidth]{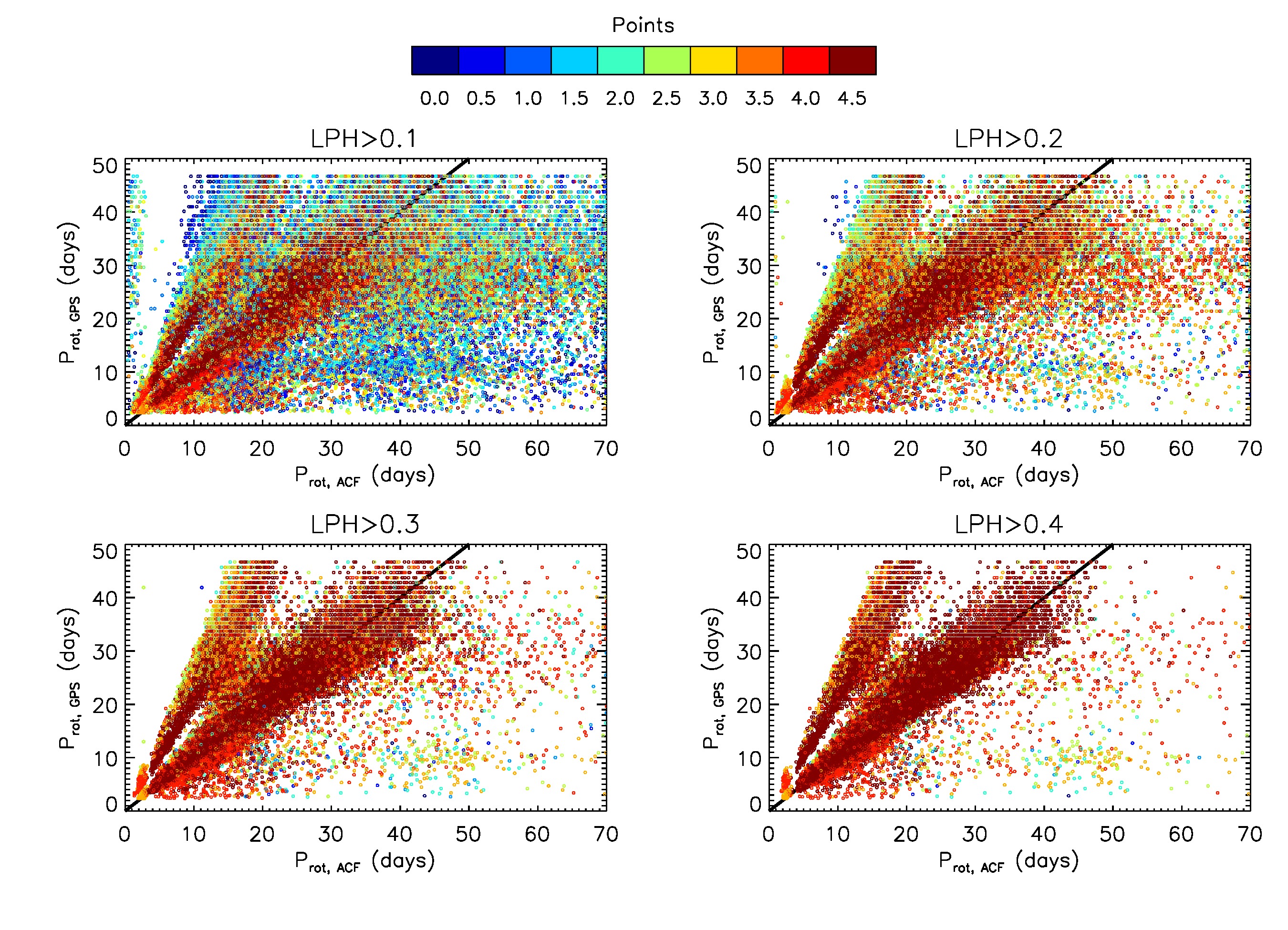}
  \caption{ACF vs. GPS periods for different LPH thresholds indicated at the top of each panel. The data are color-coded with the points system. The black line shows the 1:1 identity.}
  \label{Prot_ACF_GPS_LPH_all} 
\end{figure*}

In Fig.~\ref{Prot_ACF_GPS_LPH_all}, we show the same as in Fig.~\ref{Prot_ACF_GPS_dist} for different LPH thresholds. Additionally, the data are color-coded with the point metric defined above. It is apparent that the majority of stars with $\geq 3.0$ points are located either along the 1:1 or the 2:1 line for all LPH thresholds. This tendency becomes even more evident with increasing the LPH threshold from 0.1 to 0.4 (upper left to lower right panel), which empties most of the other plot regions. Similar to Fig.~\ref{alpha_dist}, we compute $\alpha=\Pip/P_{\rm rot,\,ACF}$ for the different LPH thresholds, and fit the distributions with a Gaussian. One derives very similar mean values increasing from $\alpha=0.204\pm0.038$ ($\rm LPH>0.1$) to $\alpha=0.214\pm0.023$ ($\rm LPH>0.4$). We further note that also the ACF peak around 15 days becomes less pronounced as LPH increases.

\subsection{Dependence of $\alpha$ on stellar parameters}
\begin{figure}
  \resizebox{\hsize}{!}{\includegraphics{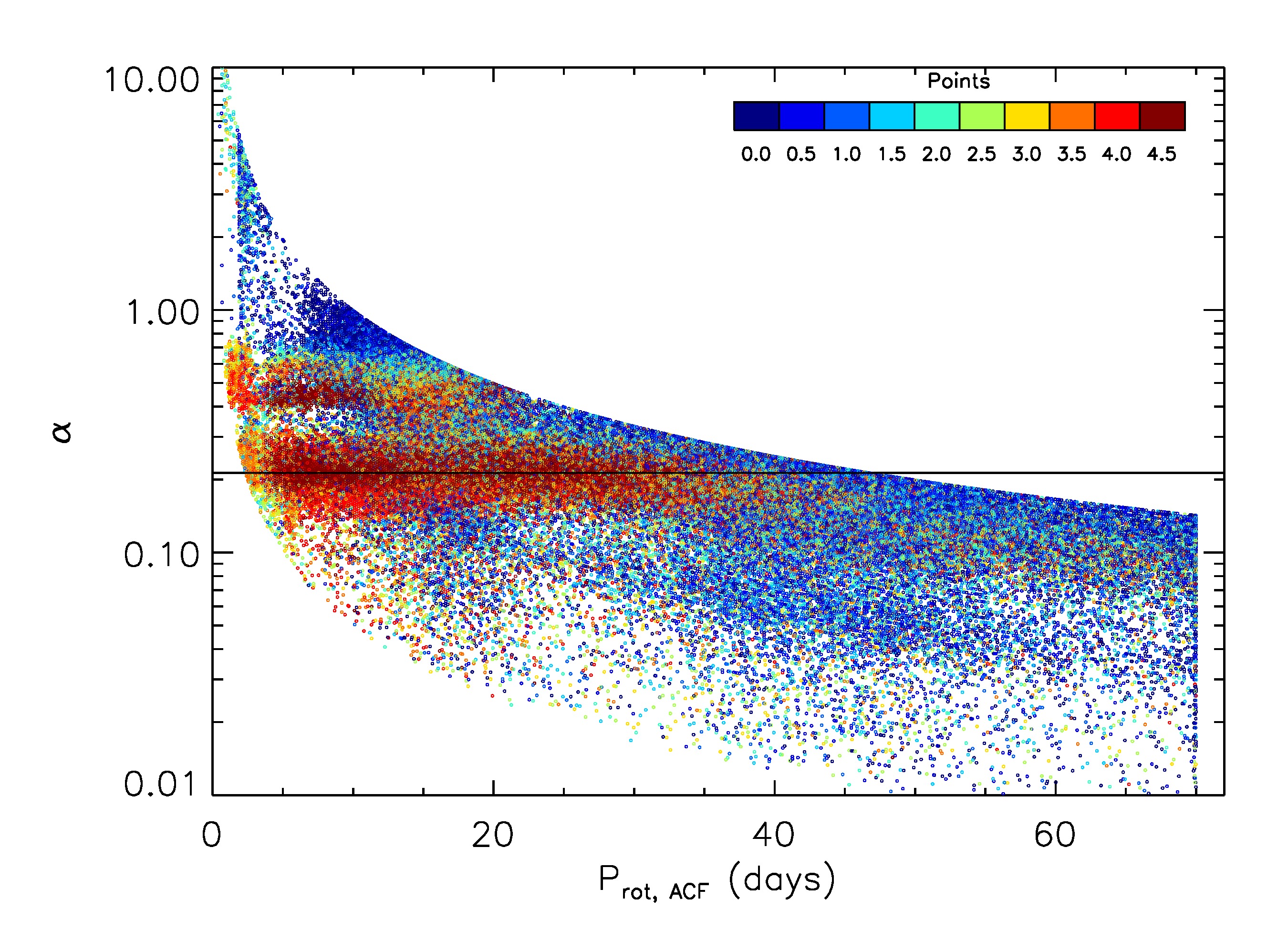}}
  \caption{ACF periods vs. $\alpha=\Pip/P_{\rm rot,\,ACF}$. The data are color-coded with the points metric. The black horizontal line indicates the mean value $\alpha=0.213$.}
  \label{Prot_alpha} 
\end{figure}

We have seen that the mean $\alpha$ values are very similar for the different samples considered so far. In this Section, we show how $\alpha$ depends on different stellar parameters, and what can be learned from that about the stars. In the following, we assume that the derived ACF periods are good measures of the stellar rotation period, at least for stars with a high number of points. In Fig.~\ref{Prot_alpha}, we show $\alpha=\Pip/P_{\rm rot,\,ACF}$ as a function of the ACF period $P_{\rm rot,\,ACF}$. We see that the stars with the largest number of points accumulate around $\alpha=0.213$ (indicated by the black horizontal line) and a bit more than twice that value (i.e. the double period branch). The plot nicely shows that $\alpha$ does not show any dependence on rotation up to periods of $\approx 35$ days. Beyond that period, there are much less stars with a high point score and, most importantly, the ACF periods become less reliable. Thus, we conclude that the same $\alpha$ value can be used to derive rotation periods of fast and slow rotators.

\begin{figure}
  \resizebox{\hsize}{!}{\includegraphics{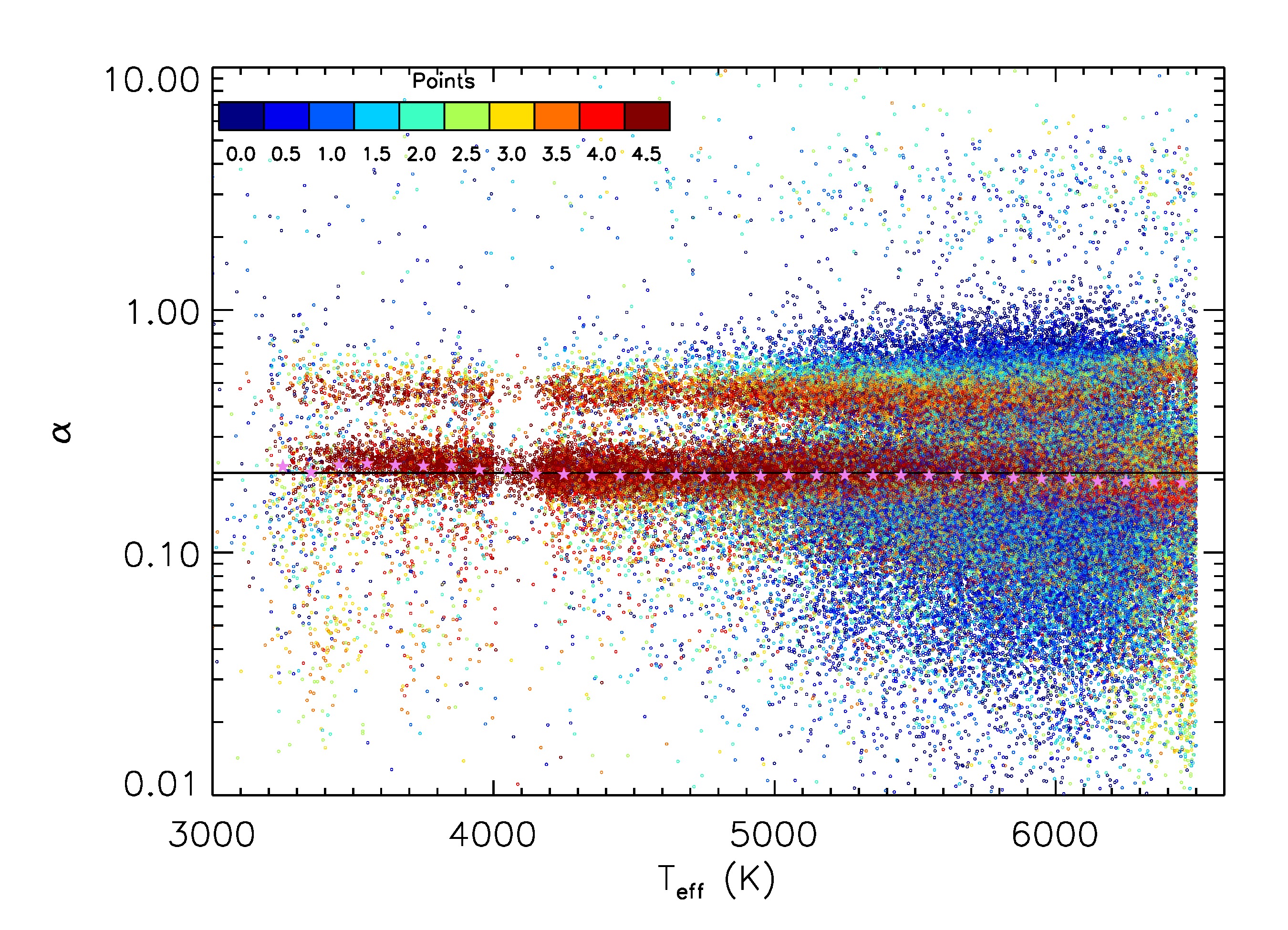}}
  \caption{Dependence of $\alpha$ on $\Teff$. The colors and the horizontal line are the same as in Fig.~\ref{Prot_alpha}. The violet star symbols show the mean $\alpha$ value in the 100\,K wide temperature bins.}
  \label{Teff_alpha} 
\end{figure}

\begin{figure}
  \resizebox{\hsize}{!}{\includegraphics{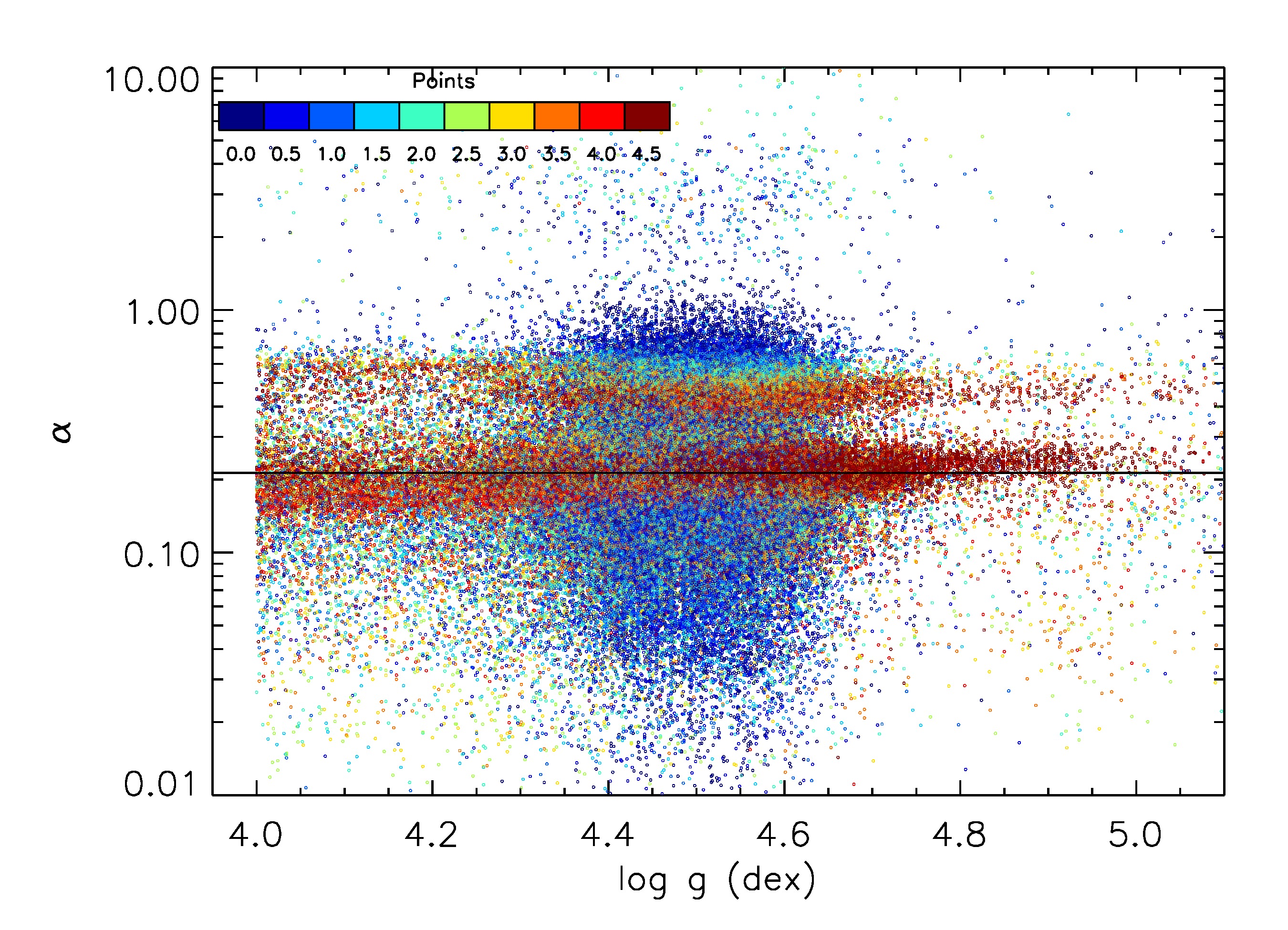}}
  \caption{Dependence of $\alpha$ on $\logg$. The colors and the horizontal line are the same as in Fig.~\ref{Prot_alpha}.}
  \label{logg_alpha} 
\end{figure}

\begin{figure}
  \resizebox{\hsize}{!}{\includegraphics{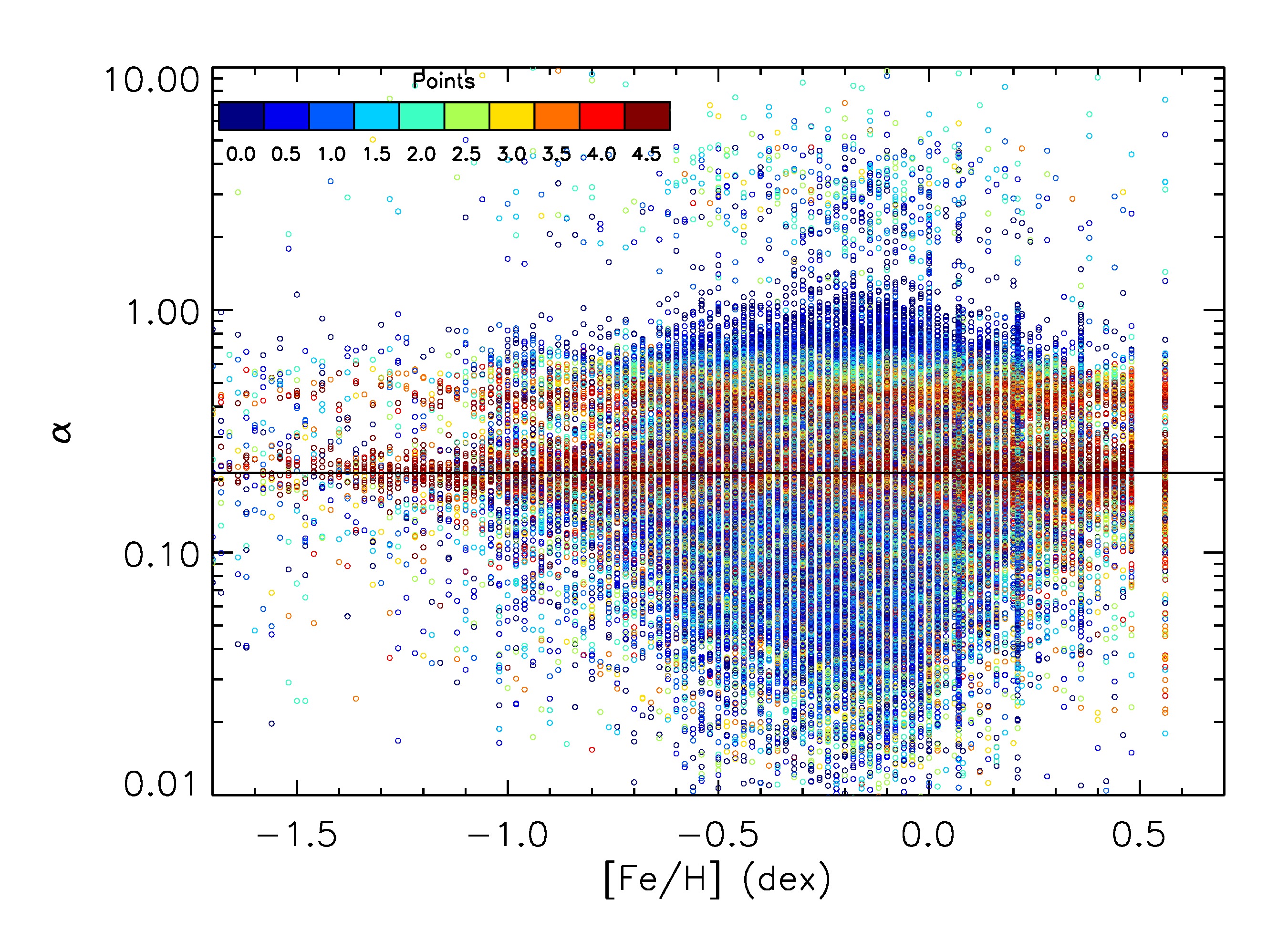}}
  \caption{Dependence of $\alpha$ on $\rm [Fe/H]$. The colors and the horizontal line are the same as in Fig.~\ref{Prot_alpha}.}
  \label{FeH_alpha} 
\end{figure}

The following three figures show the dependence of $\alpha$ on the stellar fundamental parameters effective temperature $\Teff$ (Fig.~\ref{Teff_alpha}), surface gravity $\logg$ (Fig.~\ref{logg_alpha}), and metallicity $\rm [Fe/H]$ (Fig.~\ref{FeH_alpha}). Fig.~\ref{Teff_alpha} reveals that $\alpha$ shows very little dependence on effective temperature from 4000--6000\,K. For cooler stars below 4000\,K, $\alpha$ seems to increase. The opposite effect is found for stars hotter than 6000\,K, where $\alpha$ decreases. To emphasize this effect, we restrict the main $\alpha$ branch to those stars between $0.1<\alpha<0.3$ and $\geq 3.0$ points, and overplot the mean alpha value in the 100\,K wide temperature bins as violet star symbols.

For the hot stars, the relative decrease of $\alpha$ can be explained by the fact that the inflection period is sensitive to the spot lifetimes. \citet{Giles2017} showed that spots have shorter lifetimes on G- and F-type stars compared to later-type stars. This observation was recently confirmed by \citet{Basri2022}, who used a similar approach as \citet{Giles2017} to assess the spot lifetimes. \citet{Reinhold2022} showed that the periods measured by the GPS method are shorter than the rotation period when the spot lifetimes are shorter than 2 complete rotations. As a consequence, the $\alpha$ values are smaller then the average value for these hot stars. 

For stars cooler than 4000\,K, however, we attribute the increase of $\alpha$ to another effect. \citet{Reinhold2022} further showed that the inflection point period is sensitive to the duration of a spot crossing. This dip duration depends on the spot latitude and the stellar inclination (neglecting the spot evolution on this comparatively short time scale). Spots at higher latitudes generate more sinusoidal dips in the light curves, and so have a longer dip duration than e.g. equatorial spots. The same is true for lower latitude spots on a highly-inclined star. We cannot break this degeneracy but we can argue that inclination, as a geometrical effect, is independent on effective temperature. Thus, we argue that inclination is partly responsible for the spread of $\alpha$ along the mean value but can be ruled out as explanation for the increased $\alpha$ here. 

Instead, we propose that these very cool stars exhibit spots at higher latitudes than warmer stars. This idea is tested by a simple spot model in Sect.~\ref{spot_models}. Additionally, we tested if the center-to-limb variation (CLV) for cooler stars changes such that $\alpha$ might show an increase (see appendix). However, this is not the case and can be ruled out as explanation here.

Fig.~\ref{logg_alpha} shows the dependence of $\alpha$ on $\logg$. One sees that there is also very little dependence on surface gravity, except for the high and low gravity ends. These dependencies, however, are the same as those in the previous plot because of the dependence between surface gravity and effective temperature on the main sequence. For instance, the stars at $\logg>4.8$ with the highest $\alpha$ values have almost exclusively temperatures below 4000\,K.

In Fig.~\ref{FeH_alpha} we show the dependence of $\alpha$ on metallicity. Along the main branch (black line), no dependence on metallicity is visible. This result is consistent with the latest tests of the GPS method on simulated data: \citet{Reinhold2022} found that $\alpha$ does not show any dependence on metallicity between $\rm -0.4 \leq [Fe/H]\leq 0.4$ dex for simulated time series of solar-like stars.

\subsection{Dependence of $\alpha$ on activity}
\begin{figure}
  \resizebox{\hsize}{!}{\includegraphics{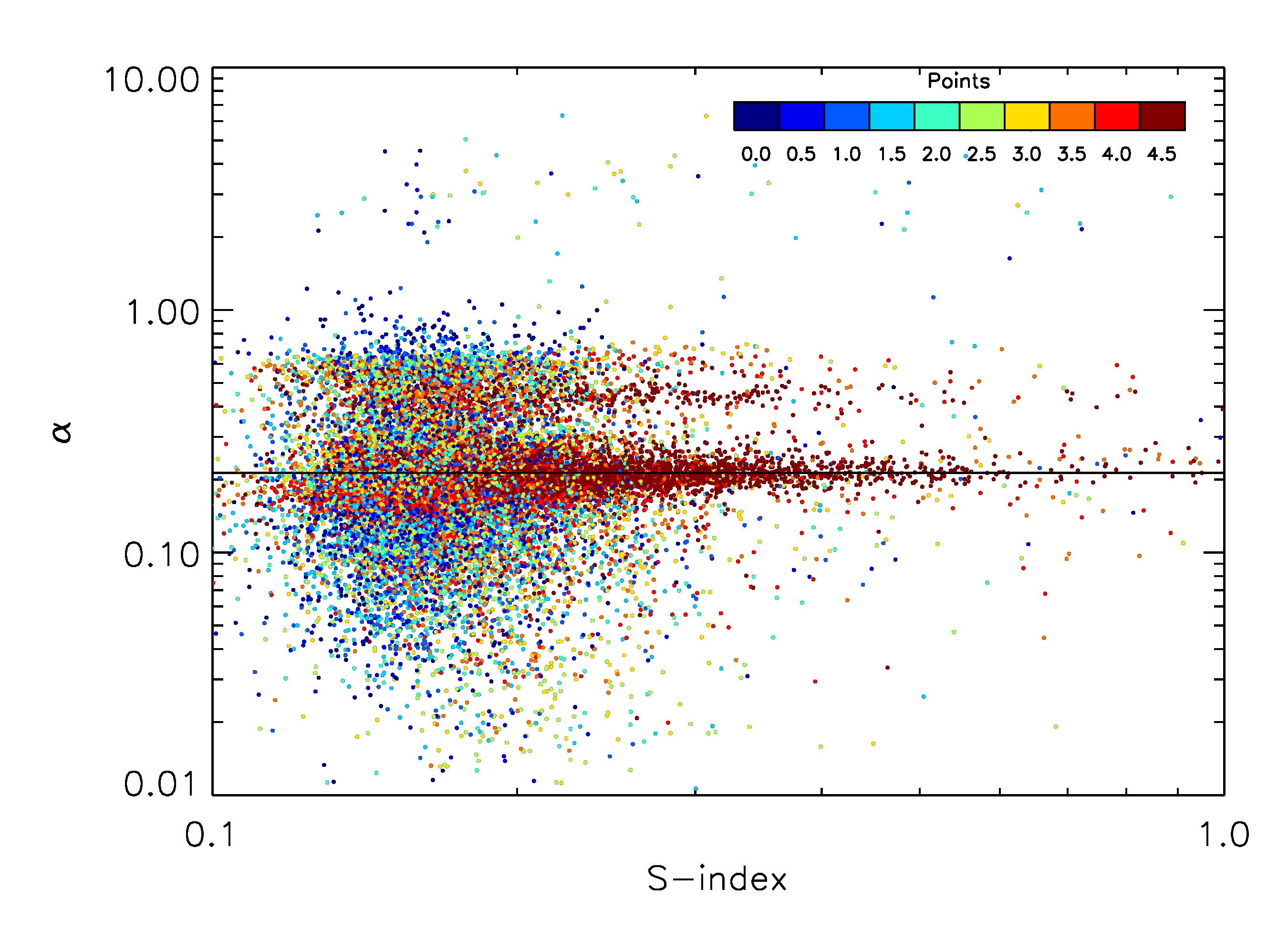}}
  \caption{Dependence of $\alpha$ on the LAMOST S-index. The colors and the horizontal line are the same as in Fig.~\ref{Prot_alpha}.}
  \label{S_ind_alpha} 
\end{figure}

In the previous section, we demonstrated that $\alpha$ shows very little dependence on rotation and the fundamental stellar parameters. Here, we test the dependence of $\alpha$ on two well-known measures of stellar activity. In Fig.~\ref{S_ind_alpha}, we show $\alpha$ as a function of the S-index, which is defined as the ratio of the flux in the Ca \ion{II} H and K lines, normalized to the flux in the R and V bands (see \citealt{Vaughan1978} for details). This activity indicator was chosen because it is a well-established measure of stellar chromospheric activity (see, e.g., \citealt{Noyes1984}). Furthermore, it is independent of the \textit{Kepler} data, in contrast to other photometric activity indices (such as the index $S_{\rm ph}$ used by \citealt{Mathur2014} or the measure $MDV$ used by \citealt{Basri2013}). The S-indices are taken from the catalog of \citet{Zhang2022} using the calibration to the Mount Wilson scale (Eq.~6 in \citealt{Zhang2022}). We find 18,797 matches between the LAMOST catalog and the stars in our sample. Fig.~\ref{S_ind_alpha} shows that $\alpha$ does not strongly depend on S-indices greater than $0.3$. However, the spread of $\alpha$ becomes much stronger toward smaller S-indices. Since the S-index itself is not corrected for its dependence on effective temperature this spread cannot solely be attributed to smaller activity.

Taking a look at the LAMOST S-index distribution shows that the vast majority of stars exhibits rather small S-indices between 0.1--0.3. This result is likely a selection effect because all these stars have effective temperatures greater than $\approx 4800$\,K. Moreover, the S-index distribution is quite narrow around the mean value of $0.2$ (even for different effective temperature bins). We attribute this to the low resolution of the LAMOST instrument, which makes it more difficult to assess the true stellar activity level, and in particular, the true dependence of $\alpha$ on activity.

\begin{figure}
  \resizebox{\hsize}{!}{\includegraphics{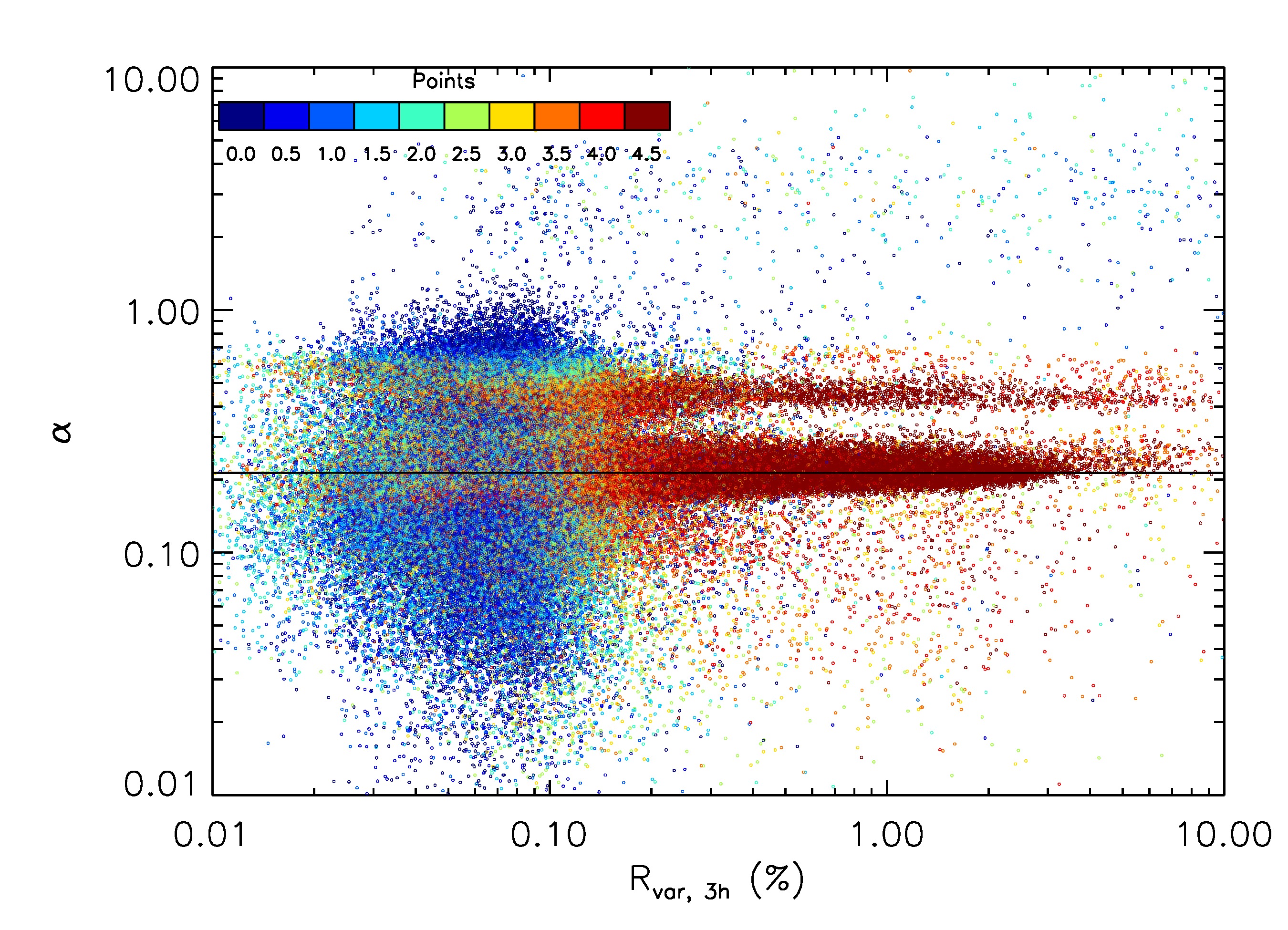}}
  \caption{Dependence of $\alpha$ on the variability range $\Rvar$. The colors and the horizontal line are the same as in Fig.~\ref{Prot_alpha}.}
  \label{Rvar_alpha} 
\end{figure}

A well-known quantity closely related to activity is the photometric variability. Here, we use the quantity $R_{var,\,3h}$ as a measure of the rotational variability, and show $\alpha$ as a function of $R_{var,\,3h}$ in Fig.~\ref{Rvar_alpha}. This figure clearly shows that $\alpha$ is almost constant down to variabilities $R_{var,\,3h}=0.2\%$, which almost equals the solar maximum variability \citep{Reinhold2020}, and that these stars have a high number of points. Down to smaller variabilities around $R_{var,\,3h}=0.1\%$, the $\alpha$ values start to show large spread. However, the horizontal branches extend down to very low variabilities with a moderate point number greater than 2 (green dots). This result clearly shows that the GPS method is able to detect the correct rotation period even for stars with very small variabilities. At the same time, also the ACF method yields the correct rotation period for these greenish stars (remember that we defined $\alpha=\Pip/P_{\rm rot,\,ACF}$). For the cloud of blue dots below $R_{var,\,3h}=0.1\%$, likely the ACF method returns a wrong period because we did not impose any LPH threshold here.

\subsection{The double period branch}\label{spot_models}
To better understand the origin of the double period branch, we employed a simple spot model. In this toy model, circular spots of a certain size and contrast can be placed on a sphere. The sphere can then be rotated, either as rigid body or differentially, and viewed from different inclination angles $i$. Since the GPS method is sensitive to the spot profile, we chose a simplified model with a single spot of fixed radius $5^\circ$ sitting at random latitudes and longitudes (both uniformly distributed) and inclination angles (uniform in $\cos i$). We arbitrarily chose a rotation period of 10 days and simulated 5 complete revolutions of rigid rotation, i.e., a time series spanning 50 days.

\begin{figure}
  \resizebox{\hsize}{!}{\includegraphics{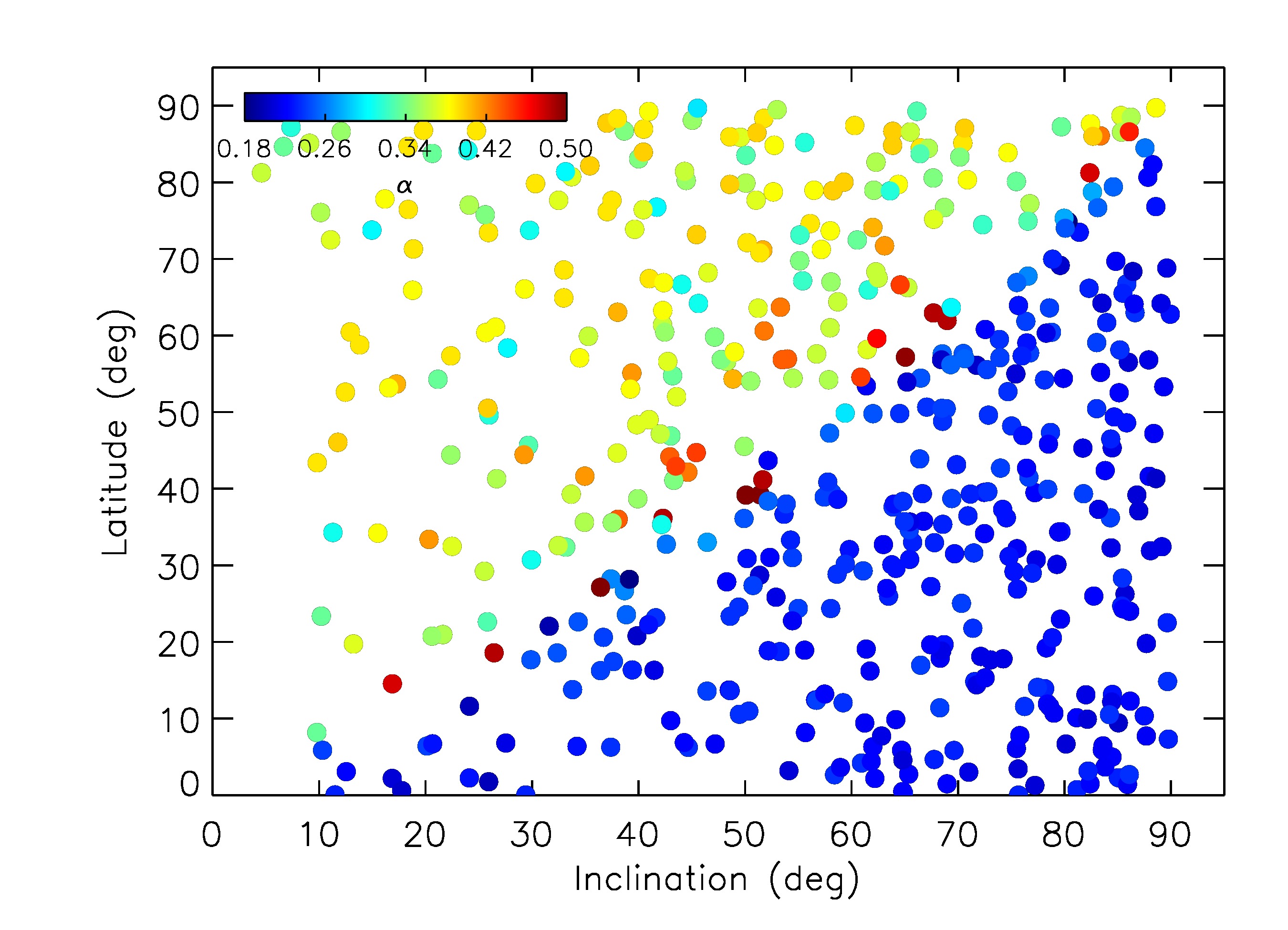}}
  \caption{GPS method applied to model light curves. The dots show the inclination and the spot latitude of the 1-spot models. The color-coding shows the derived $\alpha$ value.}
  \label{spot_model} 
\end{figure}

We computed 500 models and applied the GPS method to them. The result is shown in Fig.~\ref{spot_model}. Here, we plot the inclination of the model star against the spot latitude because both quantities affect the spot profile in the light curve, and color-code each point with the derived $\alpha$ value. There is a sharp separation between blue points in the lower right half, where the correct value of $\alpha\approx0.21$ is derived, and the yellow-greenish dots in the upper left part of the diagram, where roughly twice the correct $\alpha$ value is measured. We note that there are more dots in the lower right half, which means that we can measure the correct period for a large number of possible spot and inclination angle alignments, and that the transition between the two regimes is rather discrete (also compare Fig.~\ref{clv}).

This result again confirms that the GPS method is sensitive to the spot profile in the light curve \citep{Reinhold2022}. It is known that both higher latitude spots and/or strongly inclined stars render the spot profiles more sinusoidal, which is equivalent to an increase of the dip duration in the light curve. Since the inflection point is proportional to the dip duration, the inflection periods become larger, reaching roughly twice the correct $\alpha$ value when the dip duration equals one full rotation period.

So far, we mostly advertised the novel GPS method, especially when stellar variability is low and more irregular. However, also this method has its shortcomings, e.g., it sometimes detects twice the correct rotation period. Visual inspection of those light curves where GPS detects twice the period derived by \citetalias{McQuillan2014} or \citetalias{Santos2021} shows that this mostly happens in cases when the variability is very periodic, and so the LPH values are large. Consequently, in these cases the ACF method yields the correct period with high confidence.

To better quantify this observation, we once again consider the rotation period sample of \citetalias{Santos2021}. For this purpose, we define the number of stars in the main branch, $N_{\rm main}$, as those stars where the GPS and the \citetalias{Santos2021} period differ by less than 10\%. Similarly, the number of stars in the double period branch, $N_{\rm double}$, are defined such that the GPS and twice the \citetalias{Santos2021} period differs by less than 10\%. In the left panel of Fig.~\ref{LPH_frac}, we show the "double fraction" $N_{\rm double}/(N_{\rm main}+N_{\rm double})$ as a function of the LPH. We see that this fraction is rather flat for $\rm LPH>0.2$ (the increase in the last bin can safely be attributed to the much smaller number of stars in both branches). This so-called \textit{double floor} with a median of $\approx 6\%$ accounts for the cases with very symmetric light curves. Assuming the employed toy model correctly distinguishes between the main and double period branch, this percentage must be considered as the fraction of stars with rotation axes sufficiently inclined towards Earth that their light curves look rather sinusoidal even for spots at low latitude. 

A different behavior is seen for small LPH values where the double fraction increases. In this regime ($\rm LPH<0.2$), the light curve variability is less periodic, and the GPS method is superior to standard methods \citep{Reinhold2022}. As a consequence, here the "double period" measured by GPS is likely the correct rotation period. We also note that in this regime much fewer stars are contained in both branches because it is more difficult to detect periods in light curves with shallow periodicity (using standard methods such as \citetalias{Santos2021}). This fact is accounted for by the larger error bars assuming $\Delta N=\sqrt{N}$ for each branch.

Complementary to the LPH dependence, we show the double fraction as a function of the GPS rotation period in the right panel of Fig.~\ref{LPH_frac}. Also here different regimes can be observed: as mentioned previously, the GPS method is not sensitive to very short periods, and so the increase of the double periods for $P_{\rm rot,\,GPS}<5$ days can be ignored. Over the wide period range from 5--30 days, the double fraction shows a shallow decrease, which may be consistent with a decrease of spot latitude with rotation period but this is rather speculative. For periods greater than 30 days, a steep increase of the double period is observed (although the number of stars also strongly decreases beyond 40 days). This period range is exactly the regime where GPS is superior to classical methods and likely returns the correct period, similarly to the $\rm LPH<0.2$ range in the left panel of this figure.

\subsection{Selection of final rotation period}
\begin{figure*}
\centering
  \includegraphics[width=0.47\textwidth]{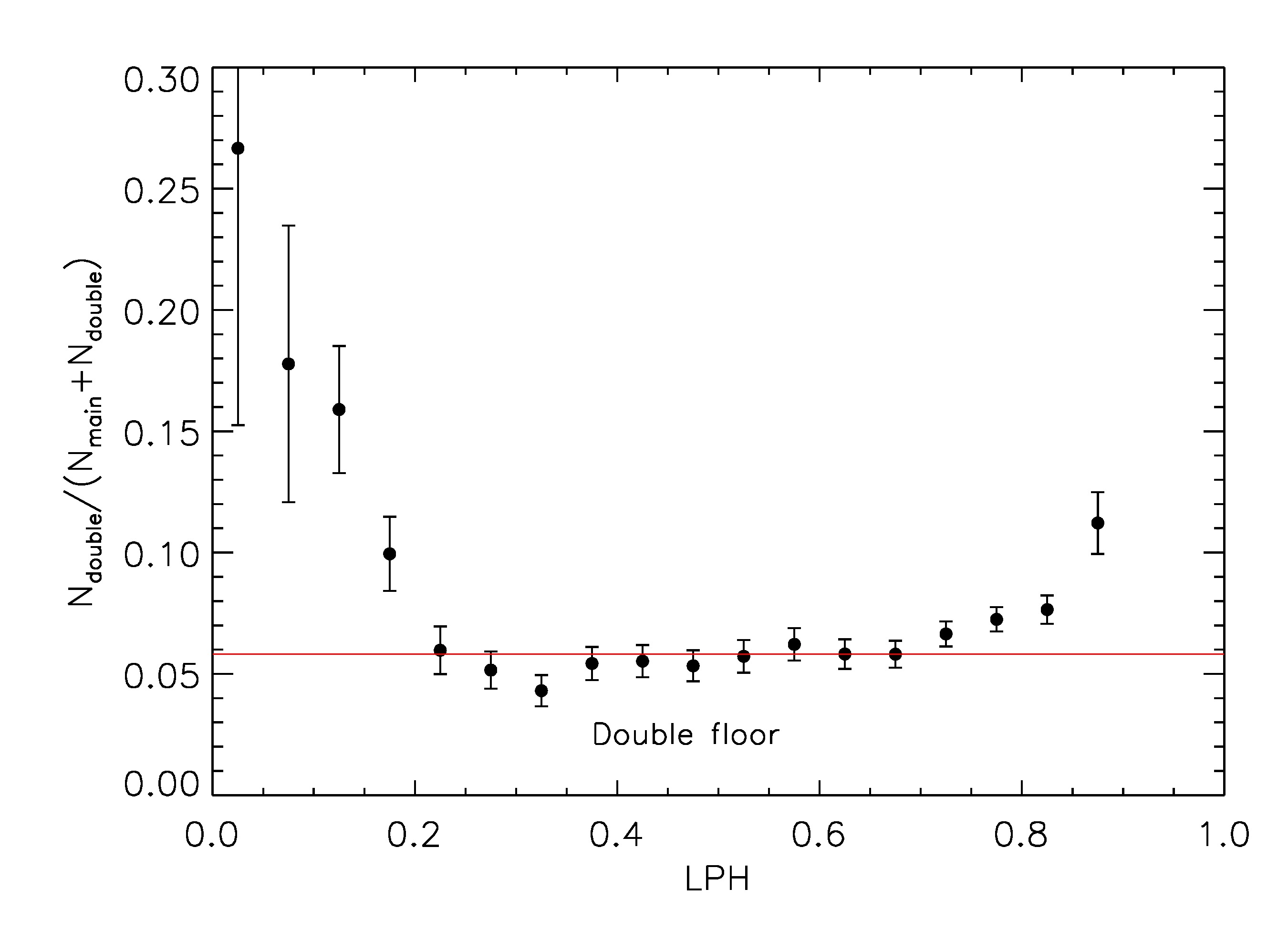}
  \includegraphics[width=0.47\textwidth]{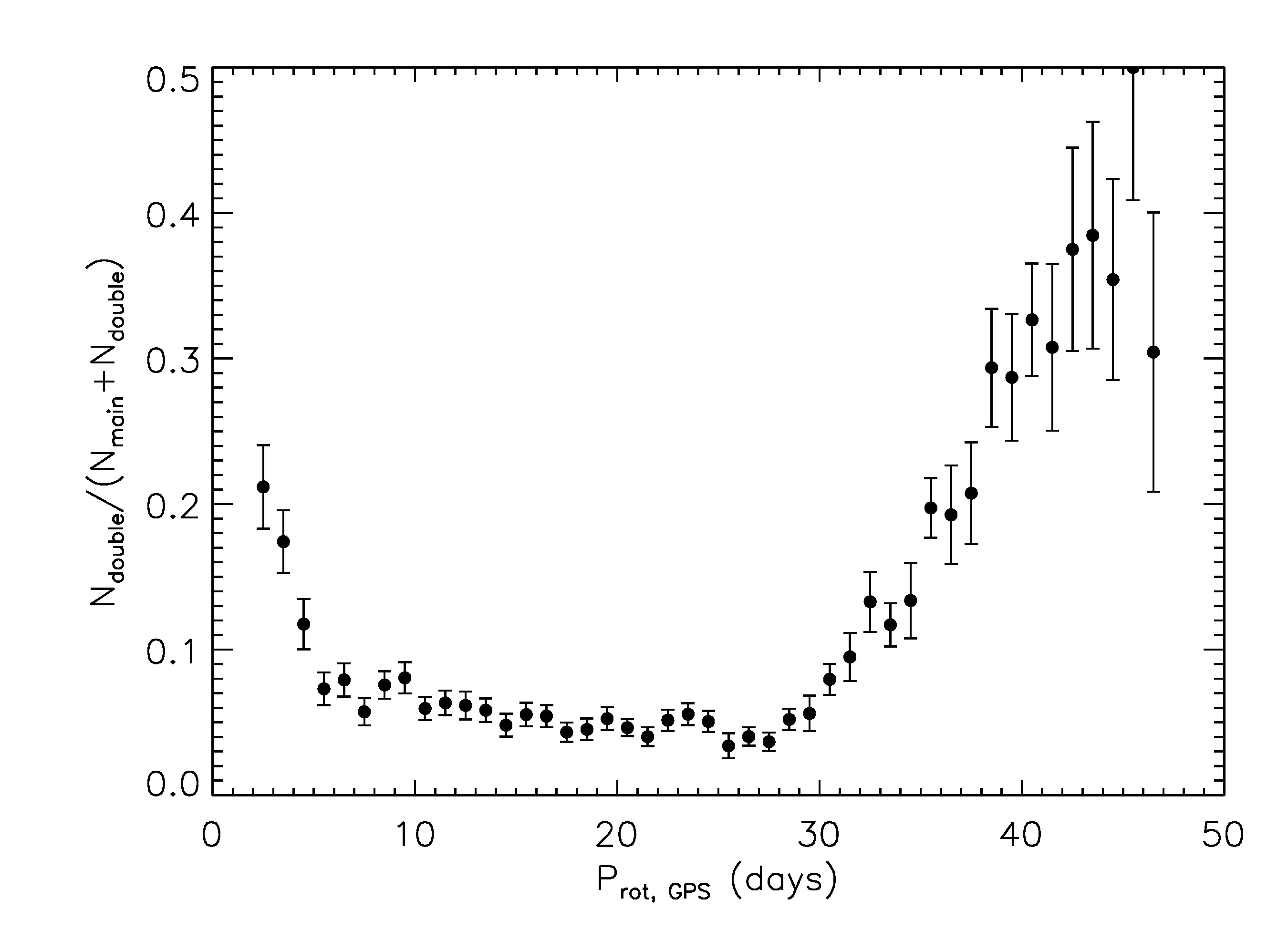}
  \caption{Fraction of stars in the double period branch as a function of the LPH (left panel) and GPS rotation period (right panel) for the stars in common with \citetalias{Santos2021}.}
  \label{LPH_frac}
\end{figure*}

Figure~\ref{LPH_frac} nicely demonstrates the region of validity of either method, and so helps to define a \textit{final} rotation period. As final rotation period $P_{\rm rot, \,fin}$, we use the ACF period if $0 < P_{\rm rot, \,ACF} \leq 10$ days and $\rm LPH \geq 0.1$. This decision is based on the fact that the ACF finds very accurate periods for fast rotators with rather sinusoidal shape and moderate to high peak heights. 

In the period range $10 < P_{\rm rot, \,ACF} \leq 20$ days, we saw that the \textit{Kepler} pipeline induces an artificial pile-up (see Fig.~\ref{Prot_ACF_GPS_dist}). We compare the ACF and GPS periods in this range, and use the ACF period if they differ by less than 10\% and the ACF fulfills the minimal requirement $\rm LPH \geq 0.1$; the GPS period is used otherwise.

For $P_{\rm rot, \,ACF} > 20$ days, the periodicity becomes weaker, the LPH values smaller, and so the GPS method superior to the ACF. Thus, we use the GPS periods here, and also for the few cases where no ACF period could be measured. The chosen parameters for the definition of the final period are summarized in Table~\ref{final_period_table}.

We note that these period thresholds are highly subjective (as most thresholds) but relies on the expertise of the authors with various kinds of light curves and frequency analysis methods. Depending on the period and LPH thresholds, the number of final rotation periods determined by the GPS and the ACF methods varies. Using our thresholds, we detect \ntot final rotation periods with 17,246 ACF and 49,917 GPS periods. We note that there exist 352 stars ($\nper-\ntot$) with $\geq 3.0$ points that have a measured GPS period but no final period assigned. These stars mostly do not satisfy the very mild $\rm LPH \geq 0.1$ thresholds. A parameter table for all stars with measured GPS period and $\geq 3.0$ points assigned is given in the appendix (Table~\ref{period_table}). 

\begin{table}
\begin{tabular}{|c|c|}
\hline
Parameter range & Method \\
\hline
$0<P_{\rm rot,\,ACF}\leq10$ days \& $\rm LPH>0.1$ & ACF \\
\hline
\makecell{$10<P_{\rm rot,\,ACF}\leq20$ days \& \\ $|P_{\rm rot,\,ACF}-P_{\rm rot,\,GPS}|/P_{\rm rot,\,ACF}<0.1$ \& $\rm LPH>0.1$} & ACF \\
\hline
\makecell{$10<P_{\rm rot,\,ACF}\leq20$ days \& \\ $|P_{\rm rot,\,ACF}-P_{\rm rot,\,GPS}|/P_{\rm rot,\,ACF}>0.1$ or $\rm 
LPH<0.1$} & GPS \\
\hline
$P_{\rm rot,\,ACF} > 20$ days & GPS \\
\hline
\end{tabular}
\caption{Summary of the final period definition.}
\label{final_period_table}
\end{table}

\section{Summary and Conclusions}

In this study we applied the novel GPS method to the light curves of tens of thousands main-sequence stars observed by the \textit{Kepler} telescope. Although this huge data set has previously been combed for rotation periods (e.g., see \citealp{Reinhold2013,McQuillan2014,Santos2021}), we showed that the GPS method was able to measure \newper periods that have not been detected before. One reason for that is that the GPS method is superior to standard frequency analysis methods (such as the ACF) for detecting rotation periods of stars with small and irregular variability.

Another important result about these \newper new periods was that their average rotation period was found to be $\sim 28$ days, and their variability $R_{\rm var, \,3h}=0.085\%$, so both the rotation period and the variability amplitude are very close to the solar values. The detection of these "solar-like" rotation periods is certainly a benefit of the GPS method.

Another advantage of the GPS method is that the high-frequency tail of the power spectrum, i.e., the period regime GPS searches for the inflection point, is much less sensitive to instrumental residuals. In stark contrast, the ACF method reveals a pile-up of periods between 10--20 days (for low to mid LPH, i.e. $0.1<LPH<0.2$). We conclude that one cannot trust ACF periods in this period and LPH range without independent confirmation by the GPS method. Another option would be to apply a customized data reduction that does not undergo a high-pass filter (likely responsible for the period pile-up at $\sim15$ days), and then apply a standard method such as ACF.

Furthermore, our work suggests a temperature dependence of the GPS calibration factor $\alpha$ for $\Teff<4000$\,K. Employing a simplified spot model, we argue that this increase of the inflection point periods is caused by spots located at higher latitude for these cool (likely early M dwarf) stars. We emphasize that this information could not be extracted by other methods before. Thus, we conclude that GPS rotation periods should be combined with other spectroscopic measurements to better constrain potential spot locations on the stellar surface.

In total, we were able to measure \ntot final rotation periods by combining the ACF and the GPS method. Compared to previous surveys, we find that 86.2\% of the \citetalias{McQuillan2014} and 77.4\% of the \citetalias{Santos2021} periods agree within 20\% with our final rotation period $P_{\rm rot,\,fin}$. For periods lower than 20 days, the difference mostly originates from the double branch, whereas for periods greater than 20 days the intrinsic uncertainty of $\alpha$ dominates. We note that the difference in this period regime is smaller for the \citetalias{McQuillan2014} sample and attribute this result to their more conservative threshold of $\rm LPH>0.3$. In the period range from 10--20 days, the GPS method is superior to classical ones (esp. the ACF) because it is not affected by the data reduction, leading to the unphysical pile-up of ACF periods around 15 days (see top panel in Fig.~\ref{Prot_ACF_GPS_dist}). We note that the ACF method usually returns more accurate periods for very periodic light curves but the shallower the periodicity gets, the more reliable the GPS period becomes, which is used in the period regime $P_{\rm rot,\,ACF}>20$ days.

In summary, this study deals with the largest set of rotation periods known to date. This work clearly demonstrated the power of novel methods (such as GPS) to detect new rotation periods even in large data sets that have been trawled through for periods before. Other promising methods might be Gaussian processes, possibly equipped with a suitable kernel function that does not a priori require periodicity. Only these methods will reveal the true rotation periods of less active stars, and so will help to improve previous solar-stellar comparison studies.

\bibliography{references_all}
\bibliographystyle{aa}

\begin{appendix}
\section{Center-to-limb variation}
We tested another possible explanation for the increase of $\alpha$ toward cooler stars, namely the center-to-limb variation (CLV). This effect describes the decline of the surface intensity from the center of the star (brightest) toward the limb (darkest). This is not only true for the quiet photosphere but also affects (in our model) cooler active region. For instance, a spot appears darker at the disc center because its contrast to the quiet photosphere is larger at the center than at the limb. In turn, the CLV might have an effect on $\alpha$ because the inflection point period is sensitive to the profile of individual spot crossing events \citep{Reinhold2022}.

The CLV has been extensively studied for the Sun (e.g., \citealt{Claret2000}). Now model grids of stars of various effective temperatures, surface gravities and metallicities are available \citep{Kostogryz2022} that are computed using the state-of-the-art radiative transfer code MPS-ATLAS \citep{Witzke2021}. 

Using the simple spot model employed in Sect.~\ref{spot_models}, we computed different sets of light curves (500 realizations each) for three different combinations of quiet photosphere and spot temperatures. Here, we fixed the the surface gravity at $\log g=4.4$ and metallicity at $\rm M/H=0.0$ dex. In each of the three sets, the same inclination and spot position (longitude and latitude) have been used to focus on the temperature dependence.

\begin{figure}
  \resizebox{\hsize}{!}{\includegraphics{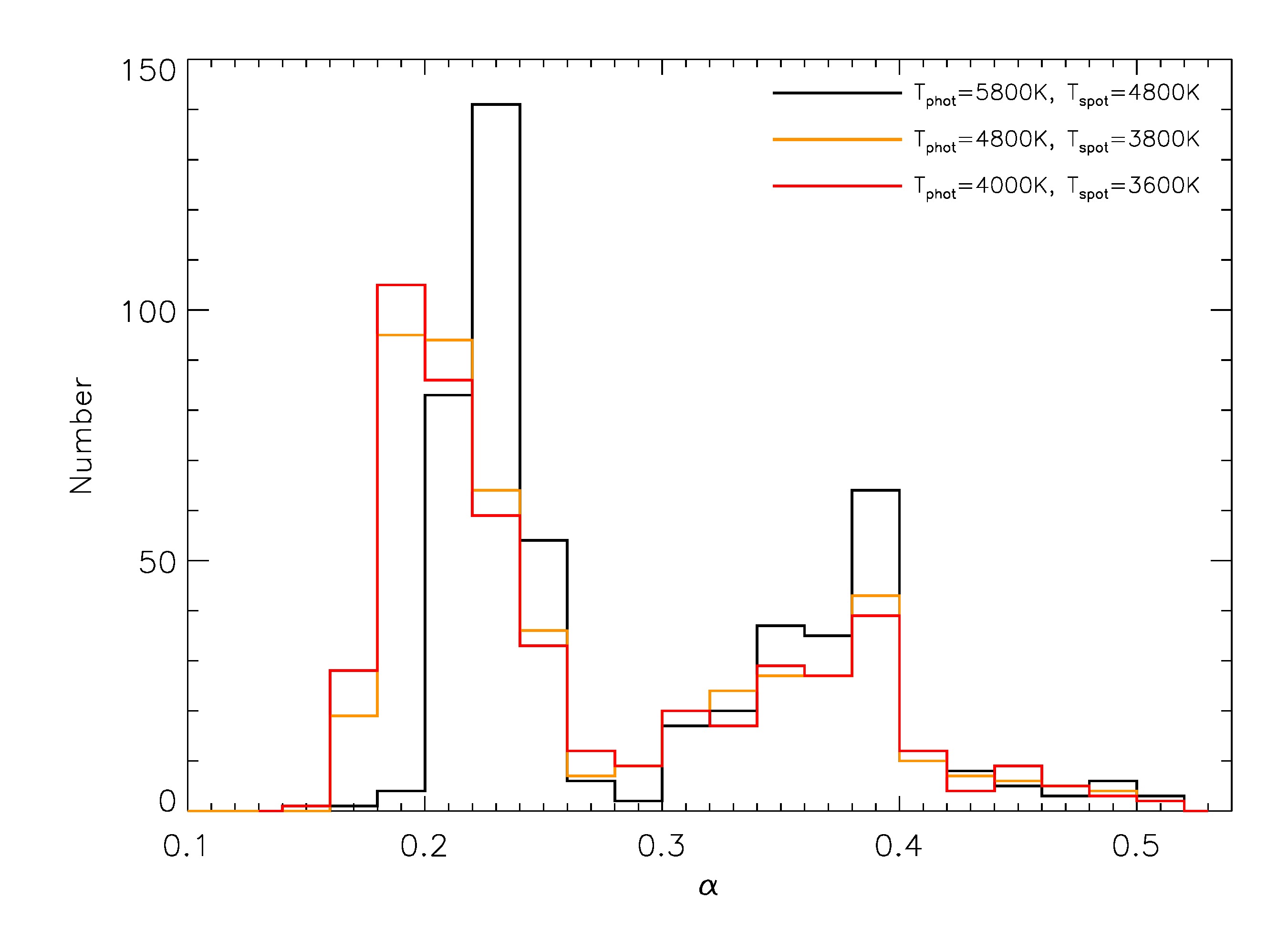}}
  \caption{GPS method applied to model light curves for three different combinations of photosphere and spot temperatures. The $\alpha$ distributions shift toward the smaller values when the temperatures decrease.}
  \label{clv} 
\end{figure}

Fig.~\ref{clv} shows the result of this small study. We chose the three combinations of photosphere and spot temperatures as $T_{\rm phot}=5800$\,K \& $T_{\rm phot}=4800$\,K (black), $T_{\rm phot}=4800$\,K \& $T_{\rm phot}=3800$\,K (orange) and $T_{\rm phot}=4000$\,K \& $T_{\rm phot}=3600$\,K (red). These combinations were chosen arbitrarily but appear reasonable since the difference between the spot and photosphere temperatures becomes smaller with decreasing effective temperature (see Fig.~7 in \citealt{Berdyugina2005}).

As expected, one can see that the majority of the distributions scatters around $\alpha=0.2$ and a smaller fraction around roughly twice that value. However, the models with smaller temperatures are shifted to smaller values of $\alpha$, in contrast to our initial assumption. Although the origin of this effect is not entirely clear, we conclude that CLV can be ruled out as an explanation for the increase of $\alpha$ toward cooler stars. The exact reason for the observed shift would require a more thorough analysis, which is beyond the scope of this study.

\section{Rotation period table}
\begin{table*}
  \centering
  \begin{tabular}{cccccccccccc}
\hline\hline
KIC & $P_{\rm rot, \, ACF}$ & LPH & $P_{\rm rot, \, GPS}$ & $h_{\rm{IP}}$ & SNR & $R_{\rm var, \, 6h}$ & points & $P_{\rm rot, \, fin}$ & method & $P_{\rm rot, \, McQ14}$ & $P_{\rm rot, \, S21}$ \\
 & (days) &  & (days) &  &  & (\%) &  & (days) &   & (days) & (days) \\
\hline
757450 & 19.30 & 0.72 & 20.95 & 1.09 & 120.4 & 1.459 & 4.5 & 19.30 & ACF & - & 19.12 \\
891901 & 23.03 & 0.47 & 15.81 & 1.06 & 44.7 & 0.093 & 4.0 & 15.81 & GPS & - & - \\
891916 & 5.52 & 0.67 & 5.59 & 1.11 & 71.2 & 1.473 & 4.5 & 5.52 & ACF & 5.52 & 5.45 \\
892376 & 13.87 & 0.60 & 17.24 & 1.10 & 208.7 & 1.060 & 4.5 & 17.24 & GPS & 1.53 & - \\
892675 & 16.31 & 0.29 & 22.35 & 1.07 & 54.7 & 0.066 & 4.5 & 22.35 & GPS & - & - \\
892834 & 13.73 & 0.87 & 16.87 & 1.10 & 264.7 & 1.259 & 4.5 & 16.87 & GPS & 13.77 & 13.61 \\
892882 & 22.08 & 0.70 & 25.46 & 1.09 & 396.5 & 0.740 & 4.5 & 25.46 & GPS & 22.29 & 21.96 \\
893033 & 26.84 & 0.65 & 27.76 & 1.10 & 392.6 & 0.604 & 4.5 & 27.76 & GPS & 27.00 & 25.94 \\
893209 & 26.13 & 0.16 & 9.81 & 1.09 & 63.0 & 0.126 & 4.0 & 9.81 & GPS & - & 4.58 \\
893286 & 29.09 & 0.40 & 27.76 & 1.09 & 207.0 & 0.424 & 4.5 & - & GPS & 28.21 & 24.54 \\
\hline
\end{tabular}

  \caption{Periods measured with the ACF and the GPS methods, including a final rotation period and a comparison to the measurements of  \citetalias{McQuillan2014} and \citetalias{Santos2021}. The full table with all \nper lines can be retrieved at the CDS.}
  \label{period_table}
\end{table*}

\end{appendix}

\end{document}